\documentclass[letterpaper,twocolumn,aps,prl,floatfix,superscriptaddress,footinbib]{revtex4-2}

\usepackage{float}	
\usepackage[caption=false]{subfig}
\usepackage{graphicx}	
\usepackage{mathtools}
\usepackage{amsfonts}
\usepackage{physics}
\usepackage{slashed}
\usepackage{placeins}
\usepackage{import}
\usepackage[colorlinks=true]{hyperref}
\usepackage[capitalise]{cleveref}
\usepackage{booktabs}
\usepackage{amsmath,amssymb}
\graphicspath{ {./figs/}}

\newcommand{\rhohat}{\hat{\rho}}
\newcommand{\ahat}{\hat{a}}
\newcommand{\Hhat}{\hat{H}}
\newcommand{\Lhat}{\hat{L}}

\begin{document}

\title{
Phase Transitions in Nonreciprocal Driven-Dissipative Condensates}

\date{\today}
\author{Ron Belyansky}
\email{ron.belyansky@gmail.com}
\affiliation{Pritzker School of Molecular Engineering, University of Chicago, Chicago, IL 60637, USA}
\author{Cheyne Weis}
\affiliation{James Franck Institute and Department of Physics,
The University of Chicago, Chicago, Illinois 60637, United States}
\author{Ryo Hanai}
\affiliation{Department of Physics, Institute of Science Tokyo, 2-12-1 Ookayama Meguro-ku, Tokyo, 152-8551, Japan}
\author{Peter B. Littlewood}
\affiliation{James Franck Institute and Department of Physics,
The University of Chicago, Chicago, Illinois 60637, United States}
\affiliation{School of Physics and Astronomy, University of St Andrews, N Haugh, St Andrews KY16 9SS, United Kingdom
}
\author{Aashish A. Clerk}
\affiliation{Pritzker School of Molecular Engineering, University of Chicago, Chicago, IL 60637, USA}
\begin{abstract}
We investigate the influence of boundaries and spatial nonreciprocity on nonequilibrium driven-dissipative phase transitions.
We focus on a one-dimensional lattice of nonlinear bosons described by a Lindblad master equation, where the interplay between coherent and incoherent dynamics generates nonreciprocal interactions between sites. Using a mean-field approach, we analyze the phase diagram under both periodic and open boundary conditions. For periodic boundaries, the system always forms a condensate at nonzero momentum and frequency, resulting in a time-dependent traveling wave pattern. In contrast, open boundaries reveal a far richer phase diagram, featuring multiple static and dynamical phases, as well as exotic phase transitions, including the spontaneous breaking of particle-hole symmetry
associated with a critical exceptional point
and phases with distinct bulk and edge behavior. Our model does not require post-selection and is experimentally realizable in platforms such as superconducting circuits.
\end{abstract}
\maketitle

{\it Introduction.---}%
Driven-dissipative many-body systems have emerged as a versatile platform for exploring novel nonequilibrium phenomena that are inaccessible in equilibrium. These systems exhibit a rich variety of exotic phases of matter, phase transitions, and unique dynamical behaviors \cite{siebererUniversalityDrivenOpen2023}.
Much of the interest is driven by the rise of programmable quantum simulators across various platforms, including exciton-polariton systems in semiconductors \cite{carusottoQuantumFluidsLight2013}, ultracold atomic systems \cite{mivehvarCavityQEDQuantum2021a}, photonic platforms \cite{ozawaTopologicalPhotonics2019,nohQuantumSimulationsManybody2016}, and superconducting circuits \cite{blaisCircuitQuantumElectrodynamics2021a}, where both coherent interactions and dissipation can be engineered.

Recent years have seen a surge of interest in nonreciprocally interacting many-body systems. 
Prototypical examples often consist of multi-species systems where the interactions between different species are asymmetric~\cite{avniNonreciprocalIsingModel2023,fruchartNonreciprocalPhaseTransitions2021a, zelleUniversalPhenomenologyCritical2024,huangActivePatternFormation2024, braunsNonreciprocalPatternFormation2024, youNonreciprocityGenericRoute2020,sahaScalarActiveMixtures2020,hanaiNonreciprocalFrustrationTime2024}.
Nonreciprocal interactions have also been shown to give rise to exotic nonequilibrium phases of matter, such as long-range order in two spatial dimensions \cite{Loos2023, Dadhichi2020, Pisegna2024}, 
time-crystalline order \cite{huangActivePatternFormation2024, braunsNonreciprocalPatternFormation2024, youNonreciprocityGenericRoute2020, fruchartNonreciprocalPhaseTransitions2021a, hanaiNonreciprocalFrustrationTime2024}, 
nonequilibrium boundary modes~\cite{muruganTopologicallyProtectedModes2017}, and chaotic phases~\cite{parkavousiEnhancedStabilityChaotic2024}. 
These phenomena have attracted significant attention in diverse fields such as active matter~\cite{bowickSymmetryThermodynamicsTopology2022}, biology~\cite{zhengTopologicalMechanismRobust2024}, and neural networks~\cite{sompolinskyTemporalAssociationAsymmetric1986}, with  explorations of analogous quantum systems also beginning to emerge \cite{khassehActiveQuantumFlocks2023a,nadolnyNonreciprocalSynchronizationActive2024,chiacchioNonreciprocalDickeModel2023b, HanaiRKKY2024, beggQuantumCriticalityOpen2024}.

Another example of nonreciprocity arises when a system exhibits preferred spatial directionality~\cite{weiderpassSolvingKineticIsing2025, godrecheDynamicsDirectedIsing2011, searaNonreciprocalInteractionsSpatially2023, veenstraNonreciprocalTopologicalSolitons2024, dasDrivenHeisenbergMagnets2002, bhattEmergentHydrodynamicsNonreciprocal2023}. A paradigmatic example is the Hatano-Nelson model \cite{hatanoVortexPinningNonHermitian1997a,hatanoLocalizationTransitionsNonHermitian1996a}, a one-dimensional system governed by a non-Hermitian Hamiltonian with asymmetric left and right hopping rates. This asymmetry causes exponential localization of all eigenstates at one boundary under open boundary conditions (OBC), while under periodic boundary conditions (PBC), the eigenstates remain extended—a phenomenon known as the non-Hermitian skin effect (NHSE) \cite{yaoEdgeStatesTopological2018a,zhangReviewNonHermitianSkin2022,okumaNonHermitianTopologicalPhenomena2023a}.
While recent studies have explored nonlinear extensions of such models \cite{ghoshHilbertSpaceFragmentation2024,manymandaSkinModesNonlinear2024,zhangSymmetryBreakingSpectral2022a,ezawaDynamicalNonlinearHigherorder2022a,zhuAnomalousSingleModeLasing2022,Brighi2024, 
veenstraNonreciprocalTopologicalSolitons2024,
beggQuantumCriticalityOpen2024,longhiModulationalInstabilityDynamical2025},
the impact on phase transitions and the specific role of boundaries, particularly beyond the framework of non-Hermitian Hamiltonians, remains largely underexplored.

\begin{figure}[b!]
	\centering
	\includegraphics[width=\linewidth]{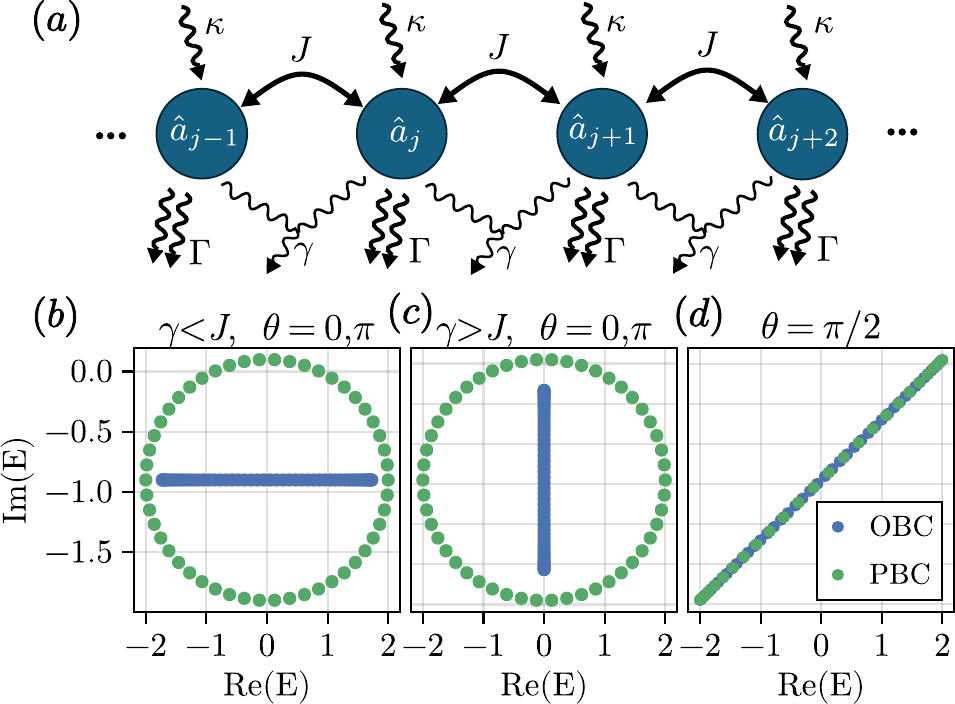}
	\caption{(a) Schematic illustration of the system. (b, c) Spectrum of the Hatano-Nelson Hamiltonian \cref{eq:Hatano-Nelson} in the regime of maximal nonreciprocity. (d) Spectrum in the reciprocal case. 
	}
	\label{fig:diag}
\end{figure}
In this work, we investigate the interplay of nonreciprocity, interactions, and boundary conditions on nonequilibrium phase transitions in a minimal driven-dissipative condensate. 
Driven dissipative condensates have been extensively studied across a range of systems \cite{kasprzakBoseEinsteinCondensation2006,amoCollectiveFluidDynamics2009,klaersBoseEinsteinCondensation2010,wertzPropagationAmplificationDynamics2012,marcosPhotonCondensationCircuit2012,carusottoQuantumFluidsLight2013,heScalingPropertiesOnedimensional2015,keelingSpontaneousRotatingVortex2008,marinoDrivenMarkovianQuantum2016a,heSpaceTimeVortexDriven2017,vercesiPhaseDiagramOnedimensional2023,fontaineKardarParisiZhang2022}. 
Recent theoretical \cite{mandalNonreciprocalTransportExciton2020} and experimental \cite{taoImaginaryGaugePotentials2025} works have begun to explore condensates in nonreciprocal settings, but the impact of nonreciprocity on phase transitions  remains unexplored.
We focus on a mean-field treatment of a one dimensional condensate 
described by a Lindblad master equation,
and find that the steady-state is highly sensitive to boundary conditions. 
While periodic boundary conditions lead to a single phase characterized by a finite momentum and frequency condensate, 
open boundaries result in a rich variety of static and dynamical phases.
These include exotic phase transitions associated with the breaking and dynamical restoration of particle-hole symmetry,
boundary-induced critical exceptional points \cite{Hanai2019polariton, Hanai2020CEP, Suchanek2023CEPentropyprod,
nadolnyNonreciprocalSynchronizationActive2024,
nakanishi2024continuoustimecrystalspt,
weisExceptionalPointsNonlinear2023,
chiacchioNonreciprocalDickeModel2023b,
youNonreciprocityGenericRoute2020, 
sahaScalarActiveMixtures2020,
zelleUniversalPhenomenologyCritical2024}, and regimes with qualitatively distinct edge and bulk dynamics.

{\it Model.---}%
We consider a chain (or a ring) of bosonic sites [see \cref{fig:diag}(a)] described by a quantum master equation \cite{breuerTheoryOpenQuantum2010}
$\partial_t \rhohat =-i\comm*{\Hhat}{\rhohat} +\sum_{\mu=1,2,3}\sum_j \mathcal{D}[\Lhat_j^{(\mu)}]\rhohat$ where the Hamiltonian, 
$\Hhat=-J\sum_j (\ahat_j^\dagger \ahat_{j+1}+\ahat_{j+1}^\dagger \ahat_j)$ describes hoppings with rate $J$, where $\hat{a}_j$ and $\hat{a}_j^\dagger$ are the site-$j$ creation and annihilation operators. The dissipators $D[\hat{L}]\hat{\rho} = \hat{L}^\dagger\hat{\rho}\hat{L} - \frac{1}{2}\{\hat{L}^\dagger\hat{L},\hat{\rho}\}$ describe three types of incoherent processes: single particle pumping with rate $\kappa$ ($\hat{L}_j^{(1)} = \sqrt{2\kappa}\hat{a}^\dagger_j$), two particle decay with rate $\Gamma$ ($\hat{L}_j^{(2)} = \sqrt{2\Gamma}\hat{a}^2_j$), and correlated single particle decay with rate $\gamma$ ($\hat{L}_j^{(3)} = \sqrt{2\gamma}(\hat{a}_j - ie^{i\theta}\hat{a}_{j+1})$).

At the quadratic level (i.e., $\Gamma=0$), this realizes the dissipative version  of the Hatano-Nelson model \cite{metelmannNonreciprocalSignalRouting2018,wanjuraTopologicalFrameworkDirectional2020a,mcdonaldNonequilibriumStationaryStates2022}. 
This can be seen, for example, from the Heisenberg equations of motion $i\partial_t \ahat_n = \sum_m (H_{\mathrm{HN}})_{nm} \ahat_m$ where the (single-particle) Hatano-Nelson Hamiltonian $H_{\mathrm{HN}}$ is
 \begin{equation}
 	\label{eq:Hatano-Nelson}
 	H_{\mathrm{HN}} = \sum_j \qty(\Delta\op{j}{j}+ J_+\op{j}{j+1}+ J_-\op{j}{j-1}),
 \end{equation} 
 where $\Delta=i(\kappa-2\gamma)$, $J_+ = -(J+\gamma e^{i\theta})$,  $J_-=-(J-\gamma e^{-i\theta})$.
The interplay between the coherent dynamics ($J$) and the correlated loss ($\gamma$) generates asymmetric hoppings rates $J_+\neq J_-$, favoring left- (right-) propagating modes when $\theta=0$ ($\theta=\pi$). The NHSE is reflected in the spectrum of \cref{eq:Hatano-Nelson}, illustrated in \cref{fig:diag}(b,c,d). 
The PBC
spectrum is an ellipse in the complex plane, whereas for 
OBC
the spectrum always falls on a line [\cref{fig:diag}(b,c)]. 
For $\theta=\pm \pi/2$, the model is reciprocal (i.e.,~$|J_+|=|J_-|$), and the ellipse shrinks into a line signaling the absence of the NHSE [\cref{fig:diag}(d)].

The quadratic system ($\Gamma=0$) can be dynamically unstable; with these interaction terms ($\Gamma\neq 0$) the system is always stable and becomes genuinely many-body.
Here, we study the mean-field dynamics of the Lindbladian, described by the equation of motion
\begin{equation}
	\label{eq:EOM}
\begin{aligned}
i\partial_t \alpha_j &= i(\kappa-2\gamma)\alpha_j -i\Gamma \abs{\alpha_{j}}^2\alpha_j \\&-(J+\gamma e^{i\theta})\alpha_{j+1} -(J-\gamma e^{-i\theta})\alpha_{j-1},
\end{aligned}
\end{equation}
where $\alpha_j\equiv \expval{\hat{a}_j}$ is the coherent state amplitude on each site. 

\FloatBarrier 
\begin{figure}[t!]
	\centering
	\includegraphics[width=\linewidth]{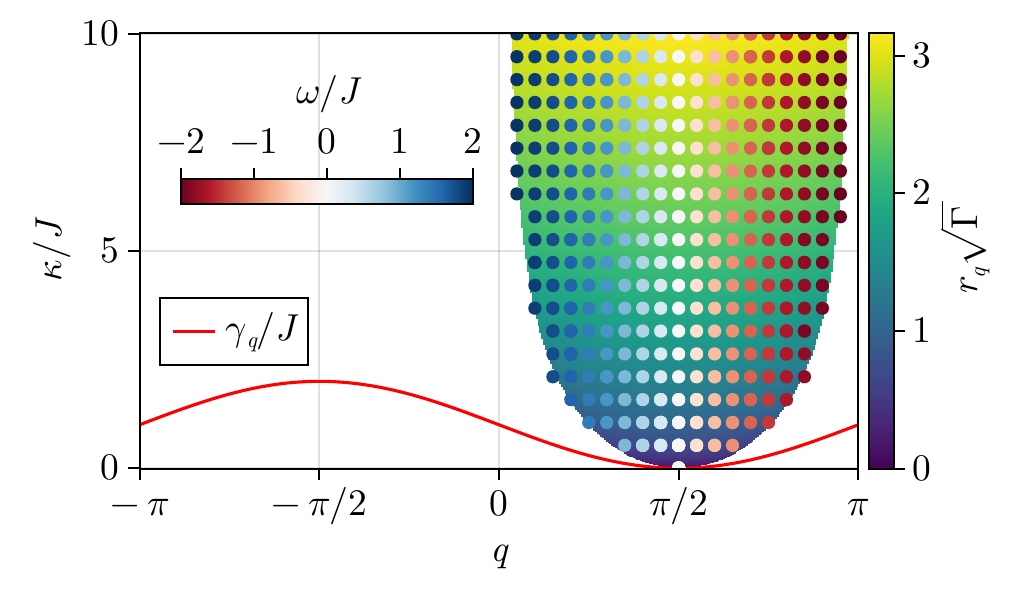}
	\caption{Stability diagram of the PBC solutions in \cref{eq:PBC-sols} for $\theta=\pi,\gamma/J=0.5$. Colored region corresponds to the stable solutions for $N\rightarrow \infty$, with the color corresponding to the amplitude $r_q$. Dots correspond to stable solutions for a finite system of $N=40$, with the color showing the oscillation frequency $\omega_q$ [\cref{eq:energy-decay}]. Red line is the momentum decay rate [\cref{eq:energy-decay}].
    }
	\label{fig:stability-pbc}
\end{figure}
{\it Periodic boundary conditions.---}%
We begin with a study of the steady-state solutions of \cref{eq:EOM} with PBC, i.e., $\alpha_{N+1}=\alpha_1$. 
For large $N$, the vacuum state $\alpha_j=0$ is unstable for any $\kappa>0$. This can be seen from the spectrum of \cref{eq:Hatano-Nelson} which is diagonalized in momentum space $H_{\mathrm{HN}} = \sum_q(i\kappa-i\gamma_q+\omega_q)\op{q}{q}$ where
\begin{equation}
	\label{eq:energy-decay}
\gamma_q = 2\gamma(1+\sin(q+\theta)) \qc \omega_q = -2J\cos(q),
\end{equation} 
are the decay rate and energy for momentum $q = 2\pi m/N,\, m=0,\cdots,N-1$. For any $\theta$, the mode at $q +\theta = -\pi/2$ (a dark mode of all $\Lhat^{(2)}_j$ dissipators) experiences no dissipation and can be excited with infinitesimal $\kappa$ (for $N\rightarrow \infty$). 

\begin{figure*}[!ht]
	\centering
	\includegraphics[width=\linewidth]{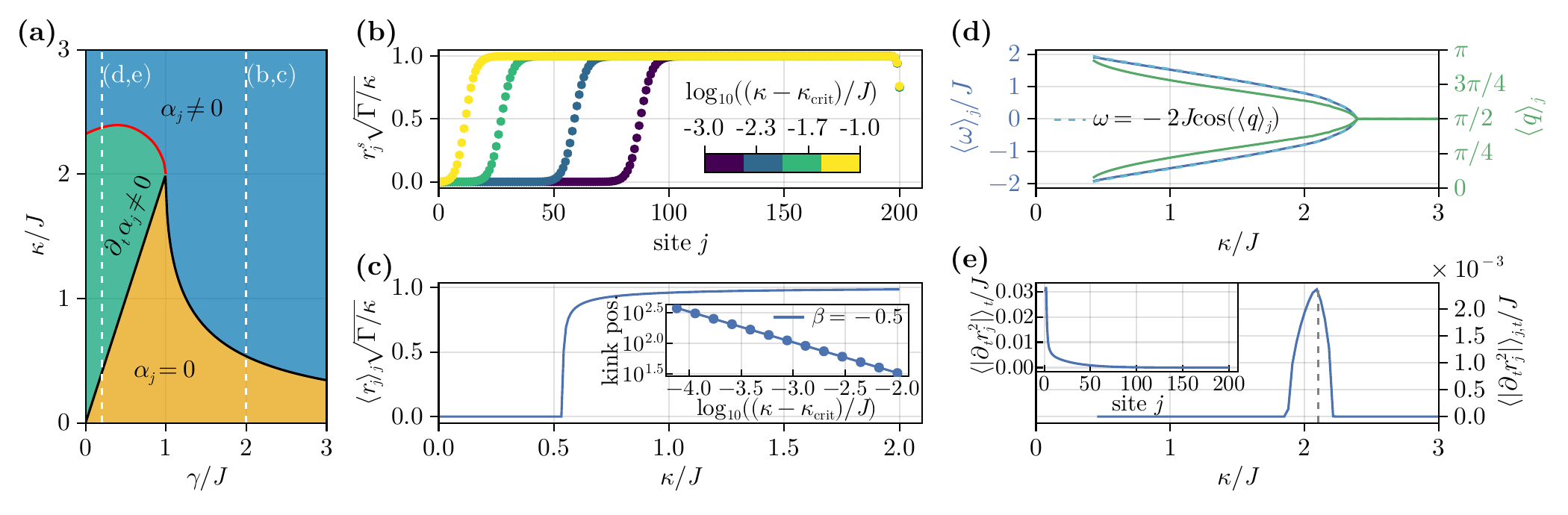}
	\caption{
		(a) Steady-state OBC phase diagram for $\theta = 0, \pi$. Dashed lines at $\gamma/J = 0.2$ and $\gamma/J = 2$ correspond to line cuts shown in panels (d, e) and (b, c), respectively. 
		(b) Spatial profile of the amplitude in the static condensate phase for different values of $\kappa$ near the phase transition for $\gamma/J = 2, \theta = \pi$. 
		(c) Average amplitude as a function of $\kappa$. The inset shows the kink position near the phase transition, with the solid line representing a power-law fit yielding a critical exponent $\beta = -0.5$. 
		(d) Average frequency (left axis, blue) and average wavevector (right axis, green) as functions of $\kappa/J$ for $\gamma/J = 0.2, \theta = \pi$. The dashed line (overlapping with the numerically obtained solid blue line) represents the PBC dispersion from \cref{eq:energy-decay}. 
		(e) Time and space average of the magnitude of the rate of change of density as a function of $\kappa$. The inset shows the spatial profile of the time-averaged density rate of change at $\kappa / J = 2.1$ (dashed line).
	}
	\label{fig:OBC_phase_diag}
\end{figure*}

Despite the nonlinearity in \cref{eq:EOM} coupling these plane waves, the time-dependent traveling wave states (i.e.~finite momentum condensates)
\begin{equation}
	\label{eq:PBC-sols}
\alpha_j(t) = r_q e^{iqj-i\omega_qt}\qc r_q = \sqrt{\frac{\kappa-\gamma_q}{\Gamma}},
\end{equation}
are solutions of \cref{eq:EOM} with PBC whenever $\gamma_q < \kappa$.  
  Not all of these plane-wave solutions are stable:  a momentum-$q$ traveling wave can become unstable
against creation of pairs at momenta $q\pm k$, for some $k$ that depends on $\kappa$ and $\gamma$ [see Supplemental Material (SM) for details \cite{supp}].
\Cref{fig:stability-pbc} shows the resulting stability diagram for $\theta=\pi$ and $\gamma/J=0.5$.
For $\kappa \rightarrow0$, only the $q^*$ mode such that $\gamma_{q^*}\rightarrow 0$ is stable ($q^*=\pi/2$ for $\theta=\pi$), as expected. 
As $\kappa\rightarrow \infty$, on the other hand, all the modes in a range of $\pi$ centered about $q^*=-\pi/2 - \theta$ are stable (corresponding to all the right-moving modes for $\theta=\pi$).   Numerically, these traveling waves appear to be the only possible behavior at long times.
For PBC and $\theta = \pi$, we thus find that the system always condenses into a single positive, finite momentum mode (chosen by the initial conditions), with a nonzero frequency $\omega_q$ (unless $q=\pi/2$). Other choices of $\theta$ lead to qualitatively similar results, with everything in \cref{fig:stability-pbc} shifted to $q^*= -\pi/2-\theta$.
Analogous traveling wave solutions also appear in the complex Landau-Ginzburg equation \cite{aransonWorldComplexGinzburgLandau2002b,ravouxStabilityAnalysisPlane2000}, the continuum version of the reciprocal $\theta = -\pi/2$ limit of our model, where the stable possible condensates are centered about $q^* = 0$.

{\it Open boundary conditions.---}%
For PBC, the angle $\theta$ (and hence the nonreciprocity $|J_+|-|J_{-}|$) does not play a significant role.  In contrast, for OBC, the value of $\theta$ is crucial to the physics.  
At $\theta=0,\pi$, the hopping is maximally nonreciprocal, and the full nonlinear system has an antilinear $\mathbb{Z}_2$ particle-hole $\mathcal{PH}$ symmetry.  At the level of  \cref{eq:EOM}, this symmetry implies that if $\alpha_j(t)$ is a solution, then so is $\mathcal{PH}[\alpha_j(t)]=e^{i\pi j}\alpha_j^*(t)$.
We will demonstrate below that this symmetry can be spontaneously broken for OBC.  At the level of the linear $\Gamma = 0$ theory, this symmetry transforms the quadratic Hatano-Nelson Hamiltonian as 
$(\mathcal{PH}) H_{\mathrm{HN}}(\mathcal{PH})^{-1}=-H_{\mathrm{HN}}$ \cite{kawabataSymmetryTopologyNonHermitian2019}.
In contrast, for more general $\theta\neq 0,\pi$, the 
$\mathcal{PH}$ symmetry is explicitly broken, making nontrivial steady states time-dependent. 
Here, we focus on the $\mathcal{PH}$-symmetric, maximally nonreciprocal case of $\theta=0,\pi$, choosing $\theta=\pi$ for concreteness. The corresponding phase diagram is shown in \cref{fig:OBC_phase_diag}(a).

The first intriguing feature of the OBC phase diagram is that the vacuum solution, $\alpha_j = 0$, becomes stable within a certain range of parameters, shown in yellow in \cref{fig:OBC_phase_diag}(a).  This is in sharp contrast to the PBC case, where the vacuum is always unstable for $\kappa>0$. The linear stability of the OBC vacuum phase is determined 
by examining the imaginary parts of the eigenvalues of \cref{eq:Hatano-Nelson} with OBC. The spectrum of \cref{eq:Hatano-Nelson} differs qualitatively depending on whether $\gamma < J$ or $\gamma > J$ [\cref{fig:diag}(b, c)], with $\gamma = J$ marking an $N$-th order exceptional point. The eigenvalues are given by \cite{supp}
\begin{equation}
	\label{eq:HN-OBC-eigs}
E_m = \begin{cases}
i(\kappa-2\gamma) +2\sqrt{J^2-\gamma^2}\cos(q_m)\qfor \gamma <J,\\
i[\kappa-2\gamma+2\sqrt{\gamma^2-J^2}\cos(q_m)]\qfor \gamma >J,
\end{cases}
\end{equation}
where $q_m = \pi m/(N+1)$, $m=1,\cdots, N$.
For $\gamma<J$, all eigenstates have the same damping rate, and the phase transition occurs at $\kappa_{\mathrm{crit}}=2\gamma$ [black line in \cref{fig:OBC_phase_diag}(a)] where all linear modes become unstable simultaneously. 
For $\gamma>J$, the $m=1$ mode ($q\rightarrow 0$ in the thermodynamic limit) is the first to become unstable, yielding the critical value $\kappa_{\mathrm{crit}} = 2\gamma -2\sqrt{\gamma^2-J^2}$.

For $\gamma > J$, as $\kappa$ is increased above $\kappa_{\mathrm{crit}}$, the system undergoes a continuous phase transition from the vacuum phase ($\expval{\alpha_j}_j = 0$) to a time-independent static condensate [blue region in \cref{fig:OBC_phase_diag}(a)] with $\expval{\alpha_j}_j \neq 0$ [\cref{fig:OBC_phase_diag}(c)], where $\expval{o}_j$ denotes the spatial average of $o$. Up to a spontaneously chosen global $U(1)$ phase, the complex amplitudes in the static phase are given by $\alpha_j^s = r_j^s e^{iq_s j}$, where $q_s = \pi/2$ ($-\pi/2$) for $\theta = \pi$ ($\theta = 0$). Interestingly, this phase order corresponds to the minimal-damping wavevector of the PBC system in \cref{eq:energy-decay}. This finite-momentum condensate emerges as $\kappa$ increases in the form of a kink that sweeps across the system from right to left [\cref{fig:OBC_phase_diag}(b)].
Numerically, we find that the height of the kink is $\sqrt{\kappa / \Gamma}$. In fact, the uniform state $\alpha_j = \sqrt{\kappa / \Gamma} e^{i\pi j/2 }$ solves \cref{eq:EOM} (with $\theta = \pi$) for all $j$ except at the edges ($j = 1, N$). As $\kappa \to \infty$, the system approaches this uniform state.

Just above the phase transition, the kink first emerges at the right edge (\( j = N \)), reflecting the localization of linear eigenstates at this boundary for \( \theta = \pi \) in the vacuum phase.
As $\kappa$ increases further, the kink shifts toward the left boundary, leading to a continuous increase in the average amplitude $\expval{r_j}_j$ [\cref{fig:OBC_phase_diag}(c)]. Numerically, we find that the kink position scales as $(\kappa - \kappa_{\mathrm{crit}})^{\beta}$ with $\beta = -0.5$ near the phase transition [inset in \cref{fig:OBC_phase_diag}(c)]. This critical exponent can be understood as follows: nonreciprocity induces directional propagation toward the right (for $\theta = \pi$). This net current balances with the local pump and loss, saturating the right-most sites at an amplitude of $\sqrt{\kappa / \Gamma}$. Disregarding these saturated sites, the remaining sites with zero amplitude form an effective smaller system.
The finite-size expression in \cref{eq:HN-OBC-eigs} can then be used to determine the new, slightly larger $\kappa$ required for the vacuum of this smaller system to become unstable.
Alternatively, solving \cref{eq:HN-OBC-eigs} for the effective system size -- or equivalently the kink position -- as a function of the distance from the phase transition ($\kappa - \kappa_{\mathrm{crit}}$) yields precisely $\beta = -0.5$.

Next, we examine the green region of the phase diagram in \cref{fig:OBC_phase_diag}(a), which corresponds to time-dependent (dynamical) phases. 
This region contains multiple  attractors that
encompass a variety of phases.
We will primarily focus on the most prominent phase, which spans nearly the entire region.
For fixed $\gamma / J < 1$, dynamical phases exist only within a finite range of $\kappa$: they transition to the vacuum phase at small $\kappa$ and become unstable to the static condensate phase at large $\kappa$. 
To locate this upper phase boundary, we perform a numerical linear stability analysis of the static condensate state, yielding the red line in \cref{fig:OBC_phase_diag}(a). 
Notably, the static condensate state $\alpha_j^s$ is $\mathcal{PH}$ symmetric (i.e., $\mathcal{PH}[\alpha_j^s] = \alpha_j^s$).

Decreasing $\kappa$ along the leftmost white dashed line in \cref{fig:OBC_phase_diag}(a) towards the red line, the dynamical matrix describing fluctuations of the static condensate develops a second zero eigenvalue, whose eigenvector coalesces with the Goldstone mode (associated with the spontaneous $U(1)$ symmetry breaking). We thus identify this portion of the red line as a second-order critical exceptional point \cite{supp}.  
Below the red line of \cref{fig:OBC_phase_diag}(a),
the system spontaneously breaks the $\mathcal{PH}$ symmetry, giving rise to two bistable time-dependent traveling wave states (wavevectors $q_\pm = q_s \pm \epsilon$) analogous to the PBC solutions in \cref{eq:PBC-sols}.
The wavevectors $q_\pm$ evolves continuously as $\kappa$ is lowered, bifurcating from $q_\pm=q_s = \pi/2$ in the static phase ($\epsilon = 0$) to either $q_- = 0$ or $q_+ = \pi$ at the point where the system enters the vacuum phase ($\epsilon = \pi/2$). 
Similarly, 
as shown in Fig.~\ref{fig:OBC_phase_diag}(d),
the frequency evolves continuously from zero, bifurcating into two branches corresponding to clockwise (CW) or counterclockwise (CCW) rotations (analogous to the ``chiral" states in Ref.~\cite{fruchartNonreciprocalPhaseTransitions2021a}).  
The $\mathcal{PH}$ symmetry maps one solution onto the other, transforming a CW rotating state into a CCW one (i.e., flipping the sign of $\omega$), while simultaneously flipping the wavevector from $q_s + \epsilon$ to $q_s - \epsilon$, and vice versa. 
Remarkably, the PBC result for the frequency in \cref{eq:energy-decay} holds perfectly for these two states, using the numerically extracted average wavevector $\langle q \rangle_j$.

\begin{figure}[t!]
	\centering
	\includegraphics[width=\linewidth]{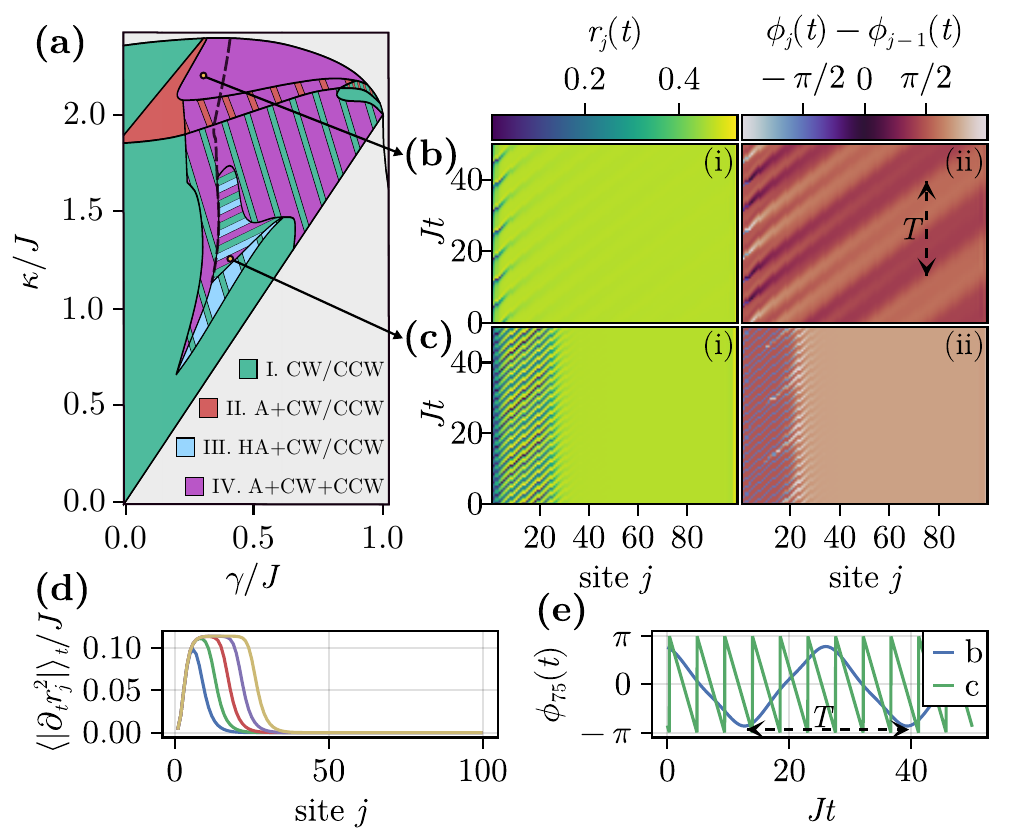}
	\caption{(a) Zoom-in on the dynamical region of the steady-state OBC phase diagram from \cref{fig:OBC_phase_diag}(a). Colored stripes indicate multistable regions with multiple phases with distinct dynamical phenomena.
    The dashed black line marks the approximate boundary where the bulk phase oscillations become chaotic (to the right of the line). (b, c) Space-time dynamics of the amplitude (i) and phase differences (ii) for representative points in phases IV and III at $\gamma=0.3, \kappa=2.2$ (b) and $\gamma=0.4, \kappa=1.25$ (c). (d) Time-averaged magnitude of the rate of change of the density in five different steady-states, corresponding to the parameters in (c). (e) Time-dependence of the phase of oscillator 75 corresponding to panels (b) and (c).}
	\label{fig:OBC_dyn_diag}
\end{figure}

{\it Edge and bulk phase transitions.---}
So far, we have considered cases where only the phase of each site varied in time. In this section, we will briefly discuss some additional dynamical phases where the amplitudes also fluctuate. These fluctuations occur predominantly at the left edge, leading to distinct bulk versus edge behavior [\cref{fig:OBC_dyn_diag}(a)].
In fact, 
\cref{fig:OBC_phase_diag}(d)
already contains such an example. 
This is illustrated in \cref{fig:OBC_phase_diag}(e), which displays the time- and space-averaged absolute value of the rate of change of the magnitude, $\langle |\partial_t r_j^2| \rangle_{j,t}$.  
This quantity becomes nonzero at $\kappa/J \sim 1.9-2.2$.
The inset in \cref{fig:OBC_phase_diag}(e) shows an example of the time-averaged spatial profile ($\expval{|\partial_t r_j^2|}_t$) of these amplitude fluctuations, revealing that they are peaked at the left edge. Away from the edge, this quantity quickly drops, and then exponentially decays into the bulk. 
Notably, these amplitude fluctuations do not affect the $\langle \omega \rangle_j$ order parameter shown in \cref{fig:OBC_phase_diag}(d), as the dominant Fourier component, even for the left-most sites, remains at the same bulk value as the average value $\langle \omega \rangle_j$ depicted in \cref{fig:OBC_phase_diag}(d).  

The dynamical region ($\partial_t\alpha_j\ne 0$) includes additional phases with distinct edge and bulk behaviors, with some regions exhibiting multistability, indicated by stripes in \cref{fig:OBC_dyn_diag}(a). Phase IV, indicated in purple in \cref{fig:OBC_dyn_diag}(a), features amplitude fluctuations at the left edge and phase-only oscillations in the bulk. A representative example ($\gamma=0.3,\kappa=2.2$) of the amplitude and phase space-time dynamics is shown in \cref{fig:OBC_dyn_diag}(b,i) and \cref{fig:OBC_dyn_diag}(b,ii), respectively.
Left of the black dashed line in \cref{fig:OBC_dyn_diag}(a), the two limit cycles of Phase I -- the CW and the CCW solutions -- merge, resulting in more complex dynamics.
The phases of the bulk sites act as pendula, involving both CW and CCW rotations [see \cref{fig:OBC_dyn_diag}(e)].
In some parts of this phase, particularly at smaller $\gamma$, the dynamics is nevertheless periodic.
Notably, the $\mathbb{Z}_2$ symmetry is dynamically restored in this case, corresponding to $\mathcal{PH}$ accompanied by the breaking of continuous time translation symmetry \cite{, ieminiBoundaryTimeCrystals2018,khemaniBriefHistoryTime2019}. 
Specifically, the $\mathcal{PH}$ operation acts as a translation by a half period, and the steady state satisfies $\alpha_j(t)=\mathcal{PH} [\alpha_j(t+T/2)]$ for some period $T$ \cite{supp}. Additional details on these phases and their symmetry classification are provided in the End Matter.

In Phase IV, the steady state alternates between exhibiting periodic and chaotic dynamics until a critical value of $\gamma$ [black dashed line in \cref{fig:OBC_dyn_diag}(a)] beyond which only the chaotic dynamics persists. 
Surprisingly, we find that despite the onset of chaos, the dynamics at the edge remain regular \cite{supp}.  
We also find a region [Phase III in \cref{fig:OBC_dyn_diag}(a)] with a hierarchy of steady states with varying edge lengths.
A representative example ($\gamma=0.4,\kappa=1.25$) is shown in \cref{fig:OBC_dyn_diag}(c). The bulk dynamics is $\mathcal{PH}$ broken, with CW or CCW phase oscillations [\cref{fig:OBC_dyn_diag}(e)] as in Phase I. We observe five steady states with progressively longer edge-amplitude fluctuations, shown in \cref{fig:OBC_dyn_diag}(d), although it remains unclear whether additional steady states exist. 
While the bulk dynamics in Phases II and III are periodic, the edge exhibits quasiperiodicity, breaking time translational symmetry and introducing an additional zero mode (beyond the existing Goldstone mode \cite{supp}). 
Exploring these edge phase transitions in greater detail is left for future work.

{\it Conclusions and outlook.---}%
We have demonstrated how the NHSE and spatial nonreciprocity can fundamentally alter the phase diagram of driven-dissipative systems in a manner that strongly depends on boundary conditions.
The remarkable differences between OBC and PBC seen here are in stark contrast to the fermionic and spin counterparts of our model~\cite{mcdonaldNonequilibriumStationaryStates2022, beggQuantumCriticalityOpen2024}, where excitations generically have finite lifetimes and decay before reaching the edge, leaving bulk properties unchanged.
In contrast, our bosonic system can possess an undamped Goldstone mode, allowing excitations to propagate to the boundaries and making boundary effects significant.

In our model, OBC result in a particularly rich phase diagram, where even the condensation phase transition exhibits unconventional features compared to its reciprocal counterpart. This intriguing difference invites further exploration to understand whether they differ in their critical behavior.
More broadly, the role of classical and quantum fluctuations beyond mean-field theory presents an exciting avenue for future research. Additionally, a deeper understanding of the observed edge phase transitions and their underlying mechanisms requires further investigation. Extending this work to higher dimensions \cite{zhuAnomalousSingleModeLasing2022} could uncover additional novel spatiotemporal patterns and exotic phases. Lastly, exploring the manifestation of these phenomena in active matter and other nonequilibrium systems remains an exciting direction for future studies.

\begin{acknowledgments}
{\it Acknowledgments.---} 
We acknowledge valuable discussions with
M. Maghrebi,
S. Diehl,
J. Binysh,
C. Corentin.  This work was supported by the Air Force Office of Scientific Research MURI program under Grant No.~FA9550-19-1-0399, the Simons Foundation through a Simons Investigator award (Grant No.~669487), 
and was completed in part with resources provided by the University of Chicago’s Research Computing Center.
RH was supported by Grant-in-Aid for Research Activity Start-up from JSPS in Japan (No. 23K19034).
CW was partially supported by NSF-MPS-PHY award 2207383.
This research benefited from Physics Frontier Center for Living Systems funded by the National Science Foundation (PHY-  2317138).
\end{acknowledgments}

\newpage
 {\it End Matter: Symmetry Classification of OBC Phases.---}%
We discuss here how each phase in our model can be elegantly classified in terms of distinct broken symmetries. There are three explicit symmetries of interest in the model.  For all values of $\theta$, Eq.~(2) of the main text admits a $U(1)\cong SO(2)$ symmetry, as well as a time-translational symmetry. For specific values of $\theta$, there is also be a discrete particle-hole symmetry, where the particle-hole operator is defined as $\mathcal{PH}[\alpha_j]=e^{i\pi j}\alpha_j^*(t)$ which is an involution.

More explicitly, when $\theta=0,\pi$, $\mathcal{PH}[\alpha_j(t)]$ obeys the same equation of motion as $\alpha_j(t)$, resulting in a $\mathbb{Z}_2$ symmetry via application of the $\mathcal{PH}$ operator. The full symmetry group for $\theta=0,\pi$ becomes $O(2)\cong SO(2)\rtimes \mathbb{Z}_2$ allowing for a richer phase diagram and collective phenomena. The continuous time-translation symmetry of the model noted above can be broken to a discrete time translation symmetry, resulting in what is often termed a time-crystalline or limit cycle phase (see e.g.~\cite{ ieminiBoundaryTimeCrystals2018,khemaniBriefHistoryTime2019}).  The remaining discrete time translational symmetry can be further broken, so that it only remains intact on a subset of measurements; this is known as \emph{quasiperiodicity}.  Time translational symmetry can also be completely broken, as is the case when one has chaotic dynamics.

\Cref{tab:symmetries} summarizes how each of the above symmetries are broken across the various condensed phases of our model. Notably, almost all of the finite number of allowed symmetry-breaking combinations are realized in this work (though a few combinations do not seem to be realized, 
e.g.~a $\mathbb{Z}_2$-symmetry broken chaotic phase).
 \begin{table*}[!t]
     \centering
     \footnotesize
     \begin{tabular}{l c c r} 
         \toprule
         Phase & $U(1)$ & $\mathbb{Z}_2$ &Continuous Time Translation \\
         \midrule
        Vacuum & Unbroken & Unbroken & Unbroken\\
         Static & Broken & Unbroken & Unbroken \\
         Phase I (Periodic Pair)& Broken & Broken & Broken to Discrete Translations \\
         Phase II, III (Quasiperiodic) & Broken & Broken & Broken (except certain measurements) \\
         Phase IV Left (Periodic Individual) & Broken & Dynamically Restored & Broken to Discrete Translations \\
         Phase IV Right (Chaotic) & Broken & Restored on Attractor &Completely Broken \\
         \bottomrule
     \end{tabular}
     \caption{Classification of OBC phases by the symmetry actions.}
     \label{tab:symmetries}
 \end{table*}

In the static phase, the solutions takes the form $\alpha_j = r_j e^{i(\phi +j\pi/2)}$ with the phase $\phi$ determined by the spontaneous breaking of $U(1)$. $\mathcal{PH}$ is left unbroken as $\alpha_j \sim\mathcal{PH}[\alpha_j]$, where we use $\sim$ to mean equivalent up to  a global rotation ($\alpha_j\to e^{i\varphi}\alpha_j$). 
In Phase I, the long-time behavior admits two branches of possible solutions with the form $\alpha_j^\pm= r_je^{i(\pm\omega t+jq_\pm + \phi)}$, where $q_\pm =\pi/2 \pm \epsilon$. The two branches of solutions are related by $\mathcal{PH}[\alpha_j^\pm]\sim\alpha_j^\mp$ implying that the $\mathcal{PH}$-symmetry is broken because there does not exist a fixed $U(1)$ transformation that allows $\mathcal{PH}[\alpha_j^\pm(t)]$ to equal $\alpha_j^\pm(t)$ for all times.

\begin{figure*}[tb]
  \centering
  \includegraphics[width=\textwidth]{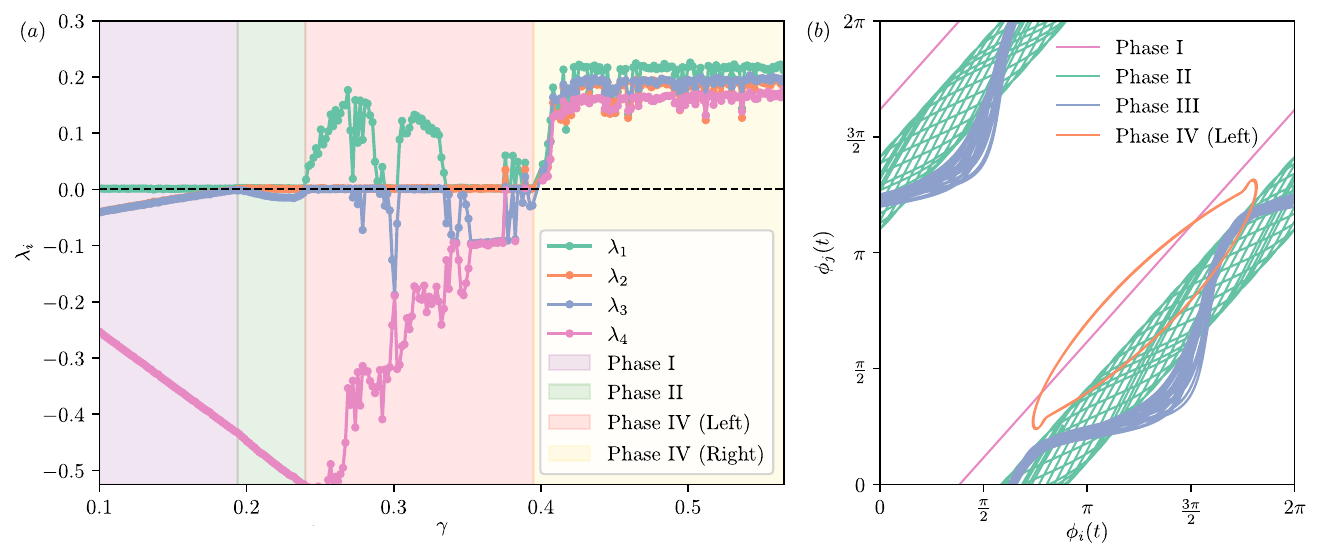}
  \caption{(a) Sweep of the 4 largest Lyapunov characteristic exponents at $\kappa=2.2$ and $N=100$. The LCEs are computed using the Benettin et al.~algorithm \cite{benettinLyapunovCharacteristicExponents1980}. Phase IV (Left) and (Right) refer to the portion of Phase IV on the corresponding side of the dotted black line in Fig.~\ref{fig:OBC_dyn_diag}(a) of the main text. (b) A phase space plot of $\phi_j(t)\coloneqq\arg(\alpha_j(t))$ for two sites near the edge of the chain with the Phase I trajectory at $\kappa=2.2,\ \gamma=0.1$ with $i=1,j=2$, Phase II at $\kappa=2,\ \gamma=0.2$ with $i=1,j=2$, Phase III at $\kappa=1.25,\ \gamma=0.4$ with $i=11,j=12$, and Phase IV at $\kappa=2.2,\ \gamma=0.35$ with $i=11,j=12$.}
  \label{fig:les}
\end{figure*}

In the remaining phases, the quasiperiodic or chaotic breaking of continuous time translational symmetry can be determined via the numerical computation of the associated Lyapunov characteristic exponents (LCEs). In Fig.~\ref{fig:les}(a), we plot the four largest LCEs for a horizontal cut through the phase diagram of Fig.~\ref{fig:OBC_dyn_diag}(a) for a fixed $\kappa = 2.2$. We divide Phase IV into two parts (left and right), depending on whether the value of $\gamma$ is to the left or right of the dashed line in Fig. \ref{fig:OBC_dyn_diag}(a).
In Phase I, there is a single zero mode (associated with the breaking of $U(1)$ symmetry), while in Phase II and Phase IV (Left) the onset of edge amplitude dynamics  coincides with the existence of two zero modes. Phase III (not shown in Fig.~\ref{fig:les}(a)) also admits two zero modes.
The existence of multiple zero modes corresponds generically to quasiperiodic dynamics due to the modes oscillating at incommensurate frequencies. 

For further insight, we plot in Fig.~\ref{fig:les}(b) phase trajectories of two sites near the left edge for various parameter regimes. 
For parameters corresponding to Phases II and III, one clearly sees non-repeating dynamics, 
whereas periodic motion is retained in trajectories corresponding to Phases I and IV (Left).
In Phases II and III, the quasiperiodicity is primarily localized near the left edge of the chain (where both the phases and amplitudes are time-dependent). Far from the edge, the dynamics becomes less distinguishable from periodic motion. The combination of amplitude and phase dynamics in Phases II and III is similar to the ``chiral+swap" phase observed in Ref.~\cite{fruchartNonreciprocalPhaseTransitions2021a} in a different model.

\begin{figure*}[tb]
  \centering
  \includegraphics[width=\textwidth]{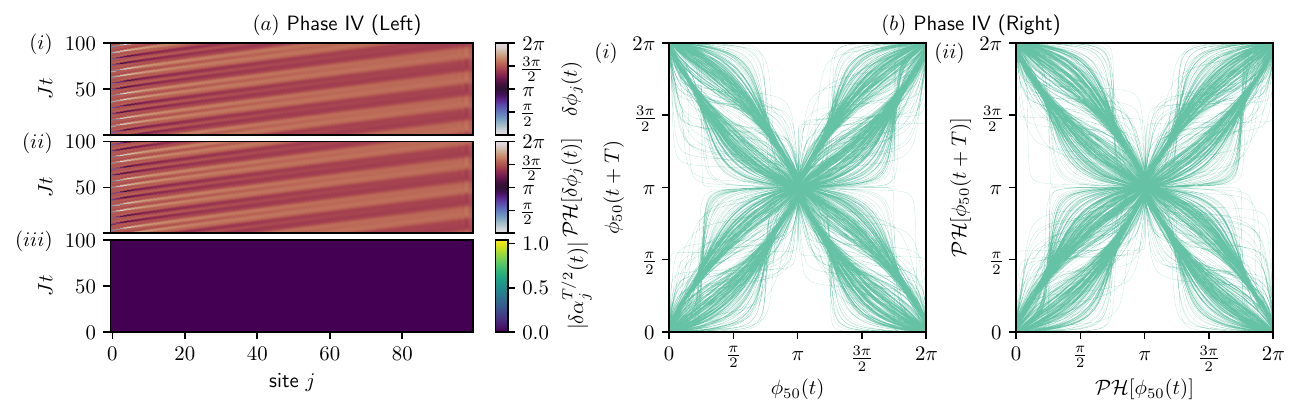}
  \caption{(a) Demonstration of the $\mathcal{PH}$ restoration at $\kappa = 2.2, \gamma = 0.3$, corresponding to point (b) in Fig.~\ref{fig:OBC_dyn_diag}(a) of the main text. 
  (a.i) Space-time dynamics of the phase differences, $\delta\phi_j(t) \coloneqq \phi_{j+1}(t) - \phi_j(t)$. 
  (a.ii) Space-time dynamics of the phase differences of the $\mathcal{PH}$-conjugated solution, i.e., $\mathcal{PH}[\delta\phi_j(t)] \coloneqq \arg(\mathcal{PH}[\alpha_{j+1}(t)]) - \arg(\mathcal{PH}[\alpha_j(t)])$. 
  (a.iii) Magnitude of $\delta\alpha_j^{T/2}(t) \coloneqq \alpha_j(t) - \mathcal{PH}[\alpha_j(t + T/2)]$, where $T \approx 26.66$ for these parameters. The global phase of $\alpha_j(t)$ is chosen so no additional rotation is needed for $\alpha_j(t) - \mathcal{PH}[\alpha_j(t+T/2)]$ to be zero. 
  (b) Time-delay embedding of the chaotic attractor at site $j=50$ for $\kappa=2.2$, $\gamma=0.5$, using phase values at times $t$ and $t+T$ with delay $T = 14/J$, comparable to the return time. 
  (i) Original attractor. (ii) $\mathcal{PH}$-conjugate attractor, with $\mathcal{PH}[\phi_j(t)] \coloneqq \arg(\mathcal{PH}[\alpha_j(t)])$.}
  \label{fig:z2restore}
\end{figure*}

Returning to Fig. \ref{fig:les}(a), note that the region of Phase IV (Left) plotted always has a single stable attractor.  However, depending on the value of $\gamma$, the phase can be chaotic (i.e.~there can be a single positive LCE).  
Phase IV (Right) generically admits multiple positive Lyapunov exponents, a situation termed in  dynamical systems theory as hyperchaos \cite{rosslerEquationHyperchaos1979,rosslerChaoticHierarchy1991}, implying that this phase admits a higher dimensional chaotic attractor than the chaos seen in Phase IV (Left).

In the nonchaotic portion of Phase IV (Left), the system retains periodicity despite the existence of two zero modes. The total phase variable $\Phi:=\sum \phi_i/N$ is static in this phase, in contrast to the $\mathcal{PH}$-broken Phase I where $\dot{\Phi}=\pm\omega$. The discrete time translation symmetry is restored because the Goldstone mode associated with $\Phi$ is now at zero frequency, making the two zero modes commensurate. A $\mathbb{Z}_2$ symmetry restoration also takes place with the explicit form $\alpha_j(t)\sim\mathcal{PH} [\alpha_j(t+T/2)]$. In Fig.~\ref{fig:z2restore}(a), we demonstrate the symmetry restoration at a periodic point of Phase IV (Left). By comparing (a.i) and (a.ii) in Fig.~\ref{fig:z2restore}, one sees that the conjugated solution is simply a time translation of the original solution. The difference between the original and the $\mathcal{PH}$-conjugate solution shifted by a half period is shown to be within numerical error of zero in Fig.~\ref{fig:z2restore}(a.iii).

In the periodic portion of Phase IV, the $\mathbb{Z}_2$ symmetry restoration $\alpha_j(t)\sim\mathcal{PH}[\alpha_j(t+T/2)]$ implies that the $\mathcal{PH}$ operation maps the entire limit cycle trajectory back onto itself. This symmetry appears to be retained as the periodic orbit bifurcates into a chaotic attractor such that the set of points on the chaotic attractor is now invariant under the $\mathcal{PH}$ operation.  This is demonstrated in 
Fig.~\ref{fig:z2restore}(b). In (b.i), we use a delay coordinate representation to give a low dimensional projection of the chaotic attractor. The attractor generated by the $\mathcal{PH}$-conjugated initial condition, shown in (b.ii), is nearly identical to the original attractor, up to small deviations due to only plotting a finite sample of the attractor. This invariance of the attractor in the chaotic phase implies that a $\mathbb{Z}_2$ symmetry remains despite the state not retaining any subgroup of the time translational symmetry.


\begin{thebibliography}{76}%
	\makeatletter
	\providecommand \@ifxundefined [1]{%
		\@ifx{#1\undefined}
	}%
	\providecommand \@ifnum [1]{%
		\ifnum #1\expandafter \@firstoftwo
		\else \expandafter \@secondoftwo
		\fi
	}%
	\providecommand \@ifx [1]{%
		\ifx #1\expandafter \@firstoftwo
		\else \expandafter \@secondoftwo
		\fi
	}%
	\providecommand \natexlab [1]{#1}%
	\providecommand \enquote  [1]{``#1''}%
	\providecommand \bibnamefont  [1]{#1}%
	\providecommand \bibfnamefont [1]{#1}%
	\providecommand \citenamefont [1]{#1}%
	\providecommand \href@noop [0]{\@secondoftwo}%
	\providecommand \href [0]{\begingroup \@sanitize@url \@href}%
	\providecommand \@href[1]{\@@startlink{#1}\@@href}%
	\providecommand \@@href[1]{\endgroup#1\@@endlink}%
	\providecommand \@sanitize@url [0]{\catcode `\\12\catcode `\$12\catcode
		`\&12\catcode `\#12\catcode `\^12\catcode `\_12\catcode `\%12\relax}%
	\providecommand \@@startlink[1]{}%
	\providecommand \@@endlink[0]{}%
	\providecommand \url  [0]{\begingroup\@sanitize@url \@url }%
	\providecommand \@url [1]{\endgroup\@href {#1}{\urlprefix }}%
	\providecommand \urlprefix  [0]{URL }%
	\providecommand \Eprint [0]{\href }%
	\providecommand \doibase [0]{https://doi.org/}%
	\providecommand \selectlanguage [0]{\@gobble}%
	\providecommand \bibinfo  [0]{\@secondoftwo}%
	\providecommand \bibfield  [0]{\@secondoftwo}%
	\providecommand \translation [1]{[#1]}%
	\providecommand \BibitemOpen [0]{}%
	\providecommand \bibitemStop [0]{}%
	\providecommand \bibitemNoStop [0]{.\EOS\space}%
	\providecommand \EOS [0]{\spacefactor3000\relax}%
	\providecommand \BibitemShut  [1]{\csname bibitem#1\endcsname}%
	\let\auto@bib@innerbib\@empty
	\bibitem [{\citenamefont {Sieberer}\ \emph {et~al.}()\citenamefont {Sieberer},
		\citenamefont {Buchhold}, \citenamefont {Marino},\ and\ \citenamefont
		{Diehl}}]{siebererUniversalityDrivenOpen2023}%
	\BibitemOpen
	\bibfield  {author} {\bibinfo {author} {\bibfnamefont {L.~M.}\ \bibnamefont
			{Sieberer}}, \bibinfo {author} {\bibfnamefont {M.}~\bibnamefont {Buchhold}},
		\bibinfo {author} {\bibfnamefont {J.}~\bibnamefont {Marino}},\ and\ \bibinfo
		{author} {\bibfnamefont {S.}~\bibnamefont {Diehl}},\ }\bibfield  {title}
	{\bibinfo {title} {Universality in driven open quantum matter},\ }\href@noop
	{} {\ }\Eprint {https://arxiv.org/abs/2312.03073} {arXiv:2312.03073}
	\BibitemShut {NoStop}%
	\bibitem [{\citenamefont {Carusotto}\ and\ \citenamefont
		{Ciuti}(2013)}]{carusottoQuantumFluidsLight2013}%
	\BibitemOpen
	\bibfield  {author} {\bibinfo {author} {\bibfnamefont {I.}~\bibnamefont
			{Carusotto}}\ and\ \bibinfo {author} {\bibfnamefont {C.}~\bibnamefont
			{Ciuti}},\ }\bibfield  {title} {\bibinfo {title} {Quantum fluids of light},\
	}\href {https://doi.org/10.1103/RevModPhys.85.299} {\bibfield  {journal}
		{\bibinfo  {journal} {Rev. Mod. Phys.}\ }\textbf {\bibinfo {volume} {85}},\
		\bibinfo {pages} {299} (\bibinfo {year} {2013})}\BibitemShut {NoStop}%
	\bibitem [{\citenamefont {Mivehvar}\ \emph {et~al.}(2021)\citenamefont
		{Mivehvar}, \citenamefont {Piazza}, \citenamefont {Donner},\ and\
		\citenamefont {Ritsch}}]{mivehvarCavityQEDQuantum2021a}%
	\BibitemOpen
	\bibfield  {author} {\bibinfo {author} {\bibfnamefont {F.}~\bibnamefont
			{Mivehvar}}, \bibinfo {author} {\bibfnamefont {F.}~\bibnamefont {Piazza}},
		\bibinfo {author} {\bibfnamefont {T.}~\bibnamefont {Donner}},\ and\ \bibinfo
		{author} {\bibfnamefont {H.}~\bibnamefont {Ritsch}},\ }\bibfield  {title}
	{\bibinfo {title} {Cavity {{QED}} with quantum gases: New paradigms in
			many-body physics},\ }\href {https://doi.org/10.1080/00018732.2021.1969727}
	{\bibfield  {journal} {\bibinfo  {journal} {Adv. Phys.}\ }\textbf {\bibinfo
			{volume} {70}},\ \bibinfo {pages} {1} (\bibinfo {year} {2021})}\BibitemShut
	{NoStop}%
	\bibitem [{\citenamefont {Ozawa}\ \emph {et~al.}(2019)\citenamefont {Ozawa},
		\citenamefont {Price}, \citenamefont {Amo}, \citenamefont {Goldman},
		\citenamefont {Hafezi}, \citenamefont {Lu}, \citenamefont {Rechtsman},
		\citenamefont {Schuster}, \citenamefont {Simon}, \citenamefont {Zilberberg},\
		and\ \citenamefont {Carusotto}}]{ozawaTopologicalPhotonics2019}%
	\BibitemOpen
	\bibfield  {author} {\bibinfo {author} {\bibfnamefont {T.}~\bibnamefont
			{Ozawa}}, \bibinfo {author} {\bibfnamefont {H.~M.}\ \bibnamefont {Price}},
		\bibinfo {author} {\bibfnamefont {A.}~\bibnamefont {Amo}}, \bibinfo {author}
		{\bibfnamefont {N.}~\bibnamefont {Goldman}}, \bibinfo {author} {\bibfnamefont
			{M.}~\bibnamefont {Hafezi}}, \bibinfo {author} {\bibfnamefont
			{L.}~\bibnamefont {Lu}}, \bibinfo {author} {\bibfnamefont {M.~C.}\
			\bibnamefont {Rechtsman}}, \bibinfo {author} {\bibfnamefont {D.}~\bibnamefont
			{Schuster}}, \bibinfo {author} {\bibfnamefont {J.}~\bibnamefont {Simon}},
		\bibinfo {author} {\bibfnamefont {O.}~\bibnamefont {Zilberberg}},\ and\
		\bibinfo {author} {\bibfnamefont {I.}~\bibnamefont {Carusotto}},\ }\bibfield
	{title} {\bibinfo {title} {Topological photonics},\ }\href
	{https://doi.org/10.1103/RevModPhys.91.015006} {\bibfield  {journal}
		{\bibinfo  {journal} {Rev. Mod. Phys.}\ }\textbf {\bibinfo {volume} {91}},\
		\bibinfo {pages} {015006} (\bibinfo {year} {2019})}\BibitemShut {NoStop}%
	\bibitem [{\citenamefont {Noh}\ and\ \citenamefont
		{Angelakis}(2016)}]{nohQuantumSimulationsManybody2016}%
	\BibitemOpen
	\bibfield  {author} {\bibinfo {author} {\bibfnamefont {C.}~\bibnamefont
			{Noh}}\ and\ \bibinfo {author} {\bibfnamefont {D.~G.}\ \bibnamefont
			{Angelakis}},\ }\bibfield  {title} {\bibinfo {title} {Quantum simulations and
			many-body physics with light},\ }\href
	{https://doi.org/10.1088/0034-4885/80/1/016401} {\bibfield  {journal}
		{\bibinfo  {journal} {Rep. Prog. Phys.}\ }\textbf {\bibinfo {volume} {80}},\
		\bibinfo {pages} {016401} (\bibinfo {year} {2016})}\BibitemShut {NoStop}%
	\bibitem [{\citenamefont {Blais}\ \emph {et~al.}(2021)\citenamefont {Blais},
		\citenamefont {Grimsmo}, \citenamefont {Girvin},\ and\ \citenamefont
		{Wallraff}}]{blaisCircuitQuantumElectrodynamics2021a}%
	\BibitemOpen
	\bibfield  {author} {\bibinfo {author} {\bibfnamefont {A.}~\bibnamefont
			{Blais}}, \bibinfo {author} {\bibfnamefont {A.~L.}\ \bibnamefont {Grimsmo}},
		\bibinfo {author} {\bibfnamefont {S.~M.}\ \bibnamefont {Girvin}},\ and\
		\bibinfo {author} {\bibfnamefont {A.}~\bibnamefont {Wallraff}},\ }\bibfield
	{title} {\bibinfo {title} {Circuit quantum electrodynamics},\ }\href
	{https://doi.org/10.1103/RevModPhys.93.025005} {\bibfield  {journal}
		{\bibinfo  {journal} {Rev. Mod. Phys.}\ }\textbf {\bibinfo {volume} {93}},\
		\bibinfo {pages} {025005} (\bibinfo {year} {2021})}\BibitemShut {NoStop}%
	\bibitem [{\citenamefont {Avni}\ \emph {et~al.}()\citenamefont {Avni},
		\citenamefont {Fruchart}, \citenamefont {Martin}, \citenamefont {Seara},\
		and\ \citenamefont {Vitelli}}]{avniNonreciprocalIsingModel2023}%
	\BibitemOpen
	\bibfield  {author} {\bibinfo {author} {\bibfnamefont {Y.}~\bibnamefont
			{Avni}}, \bibinfo {author} {\bibfnamefont {M.}~\bibnamefont {Fruchart}},
		\bibinfo {author} {\bibfnamefont {D.}~\bibnamefont {Martin}}, \bibinfo
		{author} {\bibfnamefont {D.}~\bibnamefont {Seara}},\ and\ \bibinfo {author}
		{\bibfnamefont {V.}~\bibnamefont {Vitelli}},\ }\bibfield  {title} {\bibinfo
		{title} {The non-reciprocal {{Ising}} model},\ }\href@noop {} {\ }\Eprint
	{https://arxiv.org/abs/2311.05471} {arXiv:2311.05471} \BibitemShut {NoStop}%
	\bibitem [{\citenamefont {Fruchart}\ \emph {et~al.}(2021)\citenamefont
		{Fruchart}, \citenamefont {Hanai}, \citenamefont {Littlewood},\ and\
		\citenamefont {Vitelli}}]{fruchartNonreciprocalPhaseTransitions2021a}%
	\BibitemOpen
	\bibfield  {author} {\bibinfo {author} {\bibfnamefont {M.}~\bibnamefont
			{Fruchart}}, \bibinfo {author} {\bibfnamefont {R.}~\bibnamefont {Hanai}},
		\bibinfo {author} {\bibfnamefont {P.~B.}\ \bibnamefont {Littlewood}},\ and\
		\bibinfo {author} {\bibfnamefont {V.}~\bibnamefont {Vitelli}},\ }\bibfield
	{title} {\bibinfo {title} {Non-reciprocal phase transitions},\ }\href
	{https://doi.org/10.1038/s41586-021-03375-9} {\bibfield  {journal} {\bibinfo
			{journal} {Nature}\ }\textbf {\bibinfo {volume} {592}},\ \bibinfo {pages}
		{363} (\bibinfo {year} {2021})}\BibitemShut {NoStop}%
	\bibitem [{\citenamefont {Zelle}\ \emph {et~al.}(2024)\citenamefont {Zelle},
		\citenamefont {Daviet}, \citenamefont {Rosch},\ and\ \citenamefont
		{Diehl}}]{zelleUniversalPhenomenologyCritical2024}%
	\BibitemOpen
	\bibfield  {author} {\bibinfo {author} {\bibfnamefont {C.~P.}\ \bibnamefont
			{Zelle}}, \bibinfo {author} {\bibfnamefont {R.}~\bibnamefont {Daviet}},
		\bibinfo {author} {\bibfnamefont {A.}~\bibnamefont {Rosch}},\ and\ \bibinfo
		{author} {\bibfnamefont {S.}~\bibnamefont {Diehl}},\ }\bibfield  {title}
	{\bibinfo {title} {Universal {{Phenomenology}} at {{Critical Exceptional
					Points}} of {{Nonequilibrium O}} ( {{N}} ) {{Models}}},\ }\href
	{https://doi.org/10.1103/PhysRevX.14.021052} {\bibfield  {journal} {\bibinfo
			{journal} {Phys. Rev. X}\ }\textbf {\bibinfo {volume} {14}},\ \bibinfo
		{pages} {021052} (\bibinfo {year} {2024})}\BibitemShut {NoStop}%
	\bibitem [{\citenamefont {Huang}\ \emph {et~al.}()\citenamefont {Huang},
		\citenamefont {te~Vrugt}, \citenamefont {Wittkowski},\ and\ \citenamefont
		{L{\"o}wen}}]{huangActivePatternFormation2024}%
	\BibitemOpen
	\bibfield  {author} {\bibinfo {author} {\bibfnamefont {Z.-F.}\ \bibnamefont
			{Huang}}, \bibinfo {author} {\bibfnamefont {M.}~\bibnamefont {te~Vrugt}},
		\bibinfo {author} {\bibfnamefont {R.}~\bibnamefont {Wittkowski}},\ and\
		\bibinfo {author} {\bibfnamefont {H.}~\bibnamefont {L{\"o}wen}},\ }\href@noop
	{} {\bibinfo {title} {Active pattern formation emergent from single-species
			nonreciprocity}},\ \Eprint {https://arxiv.org/abs/2404.10093}
	{arXiv:2404.10093} \BibitemShut {NoStop}%
	\bibitem [{\citenamefont {Brauns}\ and\ \citenamefont
		{Marchetti}(2024)}]{braunsNonreciprocalPatternFormation2024}%
	\BibitemOpen
	\bibfield  {author} {\bibinfo {author} {\bibfnamefont {F.}~\bibnamefont
			{Brauns}}\ and\ \bibinfo {author} {\bibfnamefont {M.~C.}\ \bibnamefont
			{Marchetti}},\ }\bibfield  {title} {\bibinfo {title} {Nonreciprocal {{Pattern
					Formation}} of {{Conserved Fields}}},\ }\href
	{https://doi.org/10.1103/PhysRevX.14.021014} {\bibfield  {journal} {\bibinfo
			{journal} {Phys. Rev. X}\ }\textbf {\bibinfo {volume} {14}},\ \bibinfo
		{pages} {021014} (\bibinfo {year} {2024})}\BibitemShut {NoStop}%
	\bibitem [{\citenamefont {You}\ \emph {et~al.}(2020)\citenamefont {You},
		\citenamefont {Baskaran},\ and\ \citenamefont
		{Marchetti}}]{youNonreciprocityGenericRoute2020}%
	\BibitemOpen
	\bibfield  {author} {\bibinfo {author} {\bibfnamefont {Z.}~\bibnamefont
			{You}}, \bibinfo {author} {\bibfnamefont {A.}~\bibnamefont {Baskaran}},\ and\
		\bibinfo {author} {\bibfnamefont {M.~C.}\ \bibnamefont {Marchetti}},\
	}\bibfield  {title} {\bibinfo {title} {Nonreciprocity as a generic route to
			traveling states},\ }\href {https://doi.org/10.1073/pnas.2010318117}
	{\bibfield  {journal} {\bibinfo  {journal} {Proc. Natl. Acad. Sci.}\ }\textbf
		{\bibinfo {volume} {117}},\ \bibinfo {pages} {19767} (\bibinfo {year}
		{2020})}\BibitemShut {NoStop}%
	\bibitem [{\citenamefont {Saha}\ \emph {et~al.}(2020)\citenamefont {Saha},
		\citenamefont {{Agudo-Canalejo}},\ and\ \citenamefont
		{Golestanian}}]{sahaScalarActiveMixtures2020}%
	\BibitemOpen
	\bibfield  {author} {\bibinfo {author} {\bibfnamefont {S.}~\bibnamefont
			{Saha}}, \bibinfo {author} {\bibfnamefont {J.}~\bibnamefont
			{{Agudo-Canalejo}}},\ and\ \bibinfo {author} {\bibfnamefont {R.}~\bibnamefont
			{Golestanian}},\ }\bibfield  {title} {\bibinfo {title} {Scalar {{Active
					Mixtures}}: {{The Nonreciprocal Cahn-Hilliard Model}}},\ }\href
	{https://doi.org/10.1103/PhysRevX.10.041009} {\bibfield  {journal} {\bibinfo
			{journal} {Phys. Rev. X}\ }\textbf {\bibinfo {volume} {10}},\ \bibinfo
		{pages} {041009} (\bibinfo {year} {2020})}\BibitemShut {NoStop}%
	\bibitem [{\citenamefont
		{Hanai}(2024)}]{hanaiNonreciprocalFrustrationTime2024}%
	\BibitemOpen
	\bibfield  {author} {\bibinfo {author} {\bibfnamefont {R.}~\bibnamefont
			{Hanai}},\ }\bibfield  {title} {\bibinfo {title} {Nonreciprocal
			{{Frustration}}: {{Time Crystalline Order-by-Disorder Phenomenon}} and a
			{{Spin-Glass-like State}}},\ }\href
	{https://doi.org/10.1103/PhysRevX.14.011029} {\bibfield  {journal} {\bibinfo
			{journal} {Phys. Rev. X}\ }\textbf {\bibinfo {volume} {14}},\ \bibinfo
		{pages} {011029} (\bibinfo {year} {2024})}\BibitemShut {NoStop}%
	\bibitem [{\citenamefont {Loos}\ \emph {et~al.}(2023)\citenamefont {Loos},
		\citenamefont {Klapp},\ and\ \citenamefont {Martynec}}]{Loos2023}%
	\BibitemOpen
	\bibfield  {author} {\bibinfo {author} {\bibfnamefont {S.~A.~M.}\
			\bibnamefont {Loos}}, \bibinfo {author} {\bibfnamefont {S.~H.~L.}\
			\bibnamefont {Klapp}},\ and\ \bibinfo {author} {\bibfnamefont
			{T.}~\bibnamefont {Martynec}},\ }\bibfield  {title} {\bibinfo {title}
		{Long-range order and directional defect propagation in the nonreciprocal
			$\mathit{XY}$ model with vision cone interactions},\ }\href
	{https://doi.org/10.1103/PhysRevLett.130.198301} {\bibfield  {journal}
		{\bibinfo  {journal} {Phys. Rev. Lett.}\ }\textbf {\bibinfo {volume} {130}},\
		\bibinfo {pages} {198301} (\bibinfo {year} {2023})}\BibitemShut {NoStop}%
	\bibitem [{\citenamefont {Dadhichi}\ \emph {et~al.}(2020)\citenamefont
		{Dadhichi}, \citenamefont {Kethapelli}, \citenamefont {Chajwa}, \citenamefont
		{Ramaswamy},\ and\ \citenamefont {Maitra}}]{Dadhichi2020}%
	\BibitemOpen
	\bibfield  {author} {\bibinfo {author} {\bibfnamefont {L.~P.}\ \bibnamefont
			{Dadhichi}}, \bibinfo {author} {\bibfnamefont {J.}~\bibnamefont
			{Kethapelli}}, \bibinfo {author} {\bibfnamefont {R.}~\bibnamefont {Chajwa}},
		\bibinfo {author} {\bibfnamefont {S.}~\bibnamefont {Ramaswamy}},\ and\
		\bibinfo {author} {\bibfnamefont {A.}~\bibnamefont {Maitra}},\ }\bibfield
	{title} {\bibinfo {title} {Nonmutual torques and the unimportance of motility
			for long-range order in two-dimensional flocks},\ }\href
	{https://doi.org/10.1103/PhysRevE.101.052601} {\bibfield  {journal} {\bibinfo
			{journal} {Phys. Rev. E}\ }\textbf {\bibinfo {volume} {101}},\ \bibinfo
		{pages} {052601} (\bibinfo {year} {2020})}\BibitemShut {NoStop}%
	\bibitem [{\citenamefont {Pisegna}\ \emph {et~al.}(2024)\citenamefont
		{Pisegna}, \citenamefont {Saha},\ and\ \citenamefont
		{Golestanian}}]{Pisegna2024}%
	\BibitemOpen
	\bibfield  {author} {\bibinfo {author} {\bibfnamefont {G.}~\bibnamefont
			{Pisegna}}, \bibinfo {author} {\bibfnamefont {S.}~\bibnamefont {Saha}},\ and\
		\bibinfo {author} {\bibfnamefont {R.}~\bibnamefont {Golestanian}},\
	}\bibfield  {title} {\bibinfo {title} {Emergent polar order in nonpolar
			mixtures with nonreciprocal interactions},\ }\href
	{https://doi.org/10.1073/pnas.2407705121} {\bibfield  {journal} {\bibinfo
			{journal} {Proc. Natl. Acad. Sci.}\ }\textbf {\bibinfo {volume} {121}},\
		\bibinfo {pages} {e2407705121} (\bibinfo {year} {2024})}\BibitemShut
	{NoStop}%
	\bibitem [{\citenamefont {Murugan}\ and\ \citenamefont
		{Vaikuntanathan}(2017)}]{muruganTopologicallyProtectedModes2017}%
	\BibitemOpen
	\bibfield  {author} {\bibinfo {author} {\bibfnamefont {A.}~\bibnamefont
			{Murugan}}\ and\ \bibinfo {author} {\bibfnamefont {S.}~\bibnamefont
			{Vaikuntanathan}},\ }\bibfield  {title} {\bibinfo {title} {Topologically
			protected modes in non-equilibrium stochastic systems},\ }\href
	{https://doi.org/10.1038/ncomms13881} {\bibfield  {journal} {\bibinfo
			{journal} {Nat Commun.}\ }\textbf {\bibinfo {volume} {8}},\ \bibinfo {pages}
		{13881} (\bibinfo {year} {2017})}\BibitemShut {NoStop}%
	\bibitem [{\citenamefont {Parkavousi}\ \emph {et~al.}()\citenamefont
		{Parkavousi}, \citenamefont {Rana}, \citenamefont {Golestanian},\ and\
		\citenamefont {Saha}}]{parkavousiEnhancedStabilityChaotic2024}%
	\BibitemOpen
	\bibfield  {author} {\bibinfo {author} {\bibfnamefont {L.}~\bibnamefont
			{Parkavousi}}, \bibinfo {author} {\bibfnamefont {N.}~\bibnamefont {Rana}},
		\bibinfo {author} {\bibfnamefont {R.}~\bibnamefont {Golestanian}},\ and\
		\bibinfo {author} {\bibfnamefont {S.}~\bibnamefont {Saha}},\ }\href@noop {}
	{\bibinfo {title} {Enhanced stability and chaotic condensates in
			multi-species non-reciprocal mixtures}},\ \Eprint
	{https://arxiv.org/abs/2408.06242} {arXiv:2408.06242} \BibitemShut {NoStop}%
	\bibitem [{\citenamefont {Bowick}\ \emph {et~al.}(2022)\citenamefont {Bowick},
		\citenamefont {Fakhri}, \citenamefont {Marchetti},\ and\ \citenamefont
		{Ramaswamy}}]{bowickSymmetryThermodynamicsTopology2022}%
	\BibitemOpen
	\bibfield  {author} {\bibinfo {author} {\bibfnamefont {M.~J.}\ \bibnamefont
			{Bowick}}, \bibinfo {author} {\bibfnamefont {N.}~\bibnamefont {Fakhri}},
		\bibinfo {author} {\bibfnamefont {M.~C.}\ \bibnamefont {Marchetti}},\ and\
		\bibinfo {author} {\bibfnamefont {S.}~\bibnamefont {Ramaswamy}},\ }\bibfield
	{title} {\bibinfo {title} {Symmetry, {{Thermodynamics}}, and {{Topology}} in
			{{Active Matter}}},\ }\href {https://doi.org/10.1103/PhysRevX.12.010501}
	{\bibfield  {journal} {\bibinfo  {journal} {Phys. Rev. X}\ }\textbf {\bibinfo
			{volume} {12}},\ \bibinfo {pages} {010501} (\bibinfo {year}
		{2022})}\BibitemShut {NoStop}%
	\bibitem [{\citenamefont {Zheng}\ and\ \citenamefont
		{Tang}(2024)}]{zhengTopologicalMechanismRobust2024}%
	\BibitemOpen
	\bibfield  {author} {\bibinfo {author} {\bibfnamefont {C.}~\bibnamefont
			{Zheng}}\ and\ \bibinfo {author} {\bibfnamefont {E.}~\bibnamefont {Tang}},\
	}\bibfield  {title} {\bibinfo {title} {A topological mechanism for robust and
			efficient global oscillations in biological networks},\ }\href
	{https://doi.org/10.1038/s41467-024-50510-x} {\bibfield  {journal} {\bibinfo
			{journal} {Nat Commun}\ }\textbf {\bibinfo {volume} {15}},\ \bibinfo {pages}
		{6453} (\bibinfo {year} {2024})}\BibitemShut {NoStop}%
	\bibitem [{\citenamefont {Sompolinsky}\ and\ \citenamefont
		{Kanter}(1986)}]{sompolinskyTemporalAssociationAsymmetric1986}%
	\BibitemOpen
	\bibfield  {author} {\bibinfo {author} {\bibfnamefont {H.}~\bibnamefont
			{Sompolinsky}}\ and\ \bibinfo {author} {\bibfnamefont {I.}~\bibnamefont
			{Kanter}},\ }\bibfield  {title} {\bibinfo {title} {Temporal {{Association}}
			in {{Asymmetric Neural Networks}}},\ }\href
	{https://doi.org/10.1103/PhysRevLett.57.2861} {\bibfield  {journal} {\bibinfo
			{journal} {Phys. Rev. Lett}\ }\textbf {\bibinfo {volume} {57}},\ \bibinfo
		{pages} {2861} (\bibinfo {year} {1986})}\BibitemShut {NoStop}%
	\bibitem [{\citenamefont {Khasseh}\ \emph {et~al.}()\citenamefont {Khasseh},
		\citenamefont {Wald}, \citenamefont {Moessner}, \citenamefont {Weber},\ and\
		\citenamefont {Heyl}}]{khassehActiveQuantumFlocks2023a}%
	\BibitemOpen
	\bibfield  {author} {\bibinfo {author} {\bibfnamefont {R.}~\bibnamefont
			{Khasseh}}, \bibinfo {author} {\bibfnamefont {S.}~\bibnamefont {Wald}},
		\bibinfo {author} {\bibfnamefont {R.}~\bibnamefont {Moessner}}, \bibinfo
		{author} {\bibfnamefont {C.~A.}\ \bibnamefont {Weber}},\ and\ \bibinfo
		{author} {\bibfnamefont {M.}~\bibnamefont {Heyl}},\ }\bibfield  {title}
	{\bibinfo {title} {Active quantum flocks},\ }\href@noop {} {\ }\Eprint
	{https://arxiv.org/abs/2308.01603} {arXiv:2308.01603} \BibitemShut {NoStop}%
	\bibitem [{\citenamefont {Nadolny}\ \emph {et~al.}(2025)\citenamefont
		{Nadolny}, \citenamefont {Bruder},\ and\ \citenamefont
		{Brunelli}}]{nadolnyNonreciprocalSynchronizationActive2024}%
	\BibitemOpen
	\bibfield  {author} {\bibinfo {author} {\bibfnamefont {T.}~\bibnamefont
			{Nadolny}}, \bibinfo {author} {\bibfnamefont {C.}~\bibnamefont {Bruder}},\
		and\ \bibinfo {author} {\bibfnamefont {M.}~\bibnamefont {Brunelli}},\
	}\bibfield  {title} {\bibinfo {title} {Nonreciprocal {{Synchronization}} of
			{{Active Quantum Spins}}},\ }\href
	{https://doi.org/10.1103/PhysRevX.15.011010} {\bibfield  {journal} {\bibinfo
			{journal} {Phys. Rev. X}\ }\textbf {\bibinfo {volume} {15}},\ \bibinfo
		{pages} {011010} (\bibinfo {year} {2025})}\BibitemShut {NoStop}%
	\bibitem [{\citenamefont {Chiacchio}\ \emph {et~al.}(2023)\citenamefont
		{Chiacchio}, \citenamefont {Nunnenkamp},\ and\ \citenamefont
		{Brunelli}}]{chiacchioNonreciprocalDickeModel2023b}%
	\BibitemOpen
	\bibfield  {author} {\bibinfo {author} {\bibfnamefont {E.~I.~R.}\
			\bibnamefont {Chiacchio}}, \bibinfo {author} {\bibfnamefont {A.}~\bibnamefont
			{Nunnenkamp}},\ and\ \bibinfo {author} {\bibfnamefont {M.}~\bibnamefont
			{Brunelli}},\ }\bibfield  {title} {\bibinfo {title} {Nonreciprocal {{Dicke
					Model}}},\ }\href {https://doi.org/10.1103/PhysRevLett.131.113602} {\bibfield
		{journal} {\bibinfo  {journal} {Phys. Rev. Lett.}\ }\textbf {\bibinfo
			{volume} {131}},\ \bibinfo {pages} {113602} (\bibinfo {year}
		{2023})}\BibitemShut {NoStop}%
	\bibitem [{\citenamefont {Hanai}\ \emph {et~al.}()\citenamefont {Hanai},
		\citenamefont {Ootsuki},\ and\ \citenamefont {Tazai}}]{HanaiRKKY2024}%
	\BibitemOpen
	\bibfield  {author} {\bibinfo {author} {\bibfnamefont {R.}~\bibnamefont
			{Hanai}}, \bibinfo {author} {\bibfnamefont {D.}~\bibnamefont {Ootsuki}},\
		and\ \bibinfo {author} {\bibfnamefont {R.}~\bibnamefont {Tazai}},\
	}\href@noop {} {\bibinfo {title} {Photoinduced non-reciprocal magnetism}},\
	\Eprint {https://arxiv.org/abs/2406.059572} {arXiv:2406.059572} \BibitemShut
	{NoStop}%
	\bibitem [{\citenamefont {Begg}\ and\ \citenamefont
		{Hanai}(2024)}]{beggQuantumCriticalityOpen2024}%
	\BibitemOpen
	\bibfield  {author} {\bibinfo {author} {\bibfnamefont {S.~E.}\ \bibnamefont
			{Begg}}\ and\ \bibinfo {author} {\bibfnamefont {R.}~\bibnamefont {Hanai}},\
	}\bibfield  {title} {\bibinfo {title} {Quantum {{Criticality}} in {{Open
					Quantum Spin Chains}} with {{Nonreciprocity}}},\ }\href
	{https://doi.org/10.1103/PhysRevLett.132.120401} {\bibfield  {journal}
		{\bibinfo  {journal} {Phys. Rev. Lett.}\ }\textbf {\bibinfo {volume} {132}},\
		\bibinfo {pages} {120401} (\bibinfo {year} {2024})}\BibitemShut {NoStop}%
	\bibitem [{\citenamefont {Weiderpass}\ \emph {et~al.}(2025)\citenamefont
		{Weiderpass}, \citenamefont {Sharma},\ and\ \citenamefont
		{Sethi}}]{weiderpassSolvingKineticIsing2025}%
	\BibitemOpen
	\bibfield  {author} {\bibinfo {author} {\bibfnamefont {G.~A.}\ \bibnamefont
			{Weiderpass}}, \bibinfo {author} {\bibfnamefont {M.}~\bibnamefont {Sharma}},\
		and\ \bibinfo {author} {\bibfnamefont {S.}~\bibnamefont {Sethi}},\ }\bibfield
	{title} {\bibinfo {title} {Solving the kinetic {{Ising}} model with
			nonreciprocity},\ }\href {https://doi.org/10.1103/PhysRevE.111.024107}
	{\bibfield  {journal} {\bibinfo  {journal} {Physical Review E}\ }\textbf
		{\bibinfo {volume} {111}},\ \bibinfo {pages} {024107} (\bibinfo {year}
		{2025})}\BibitemShut {NoStop}%
	\bibitem [{\citenamefont
		{Godr{\`e}che}(2011)}]{godrecheDynamicsDirectedIsing2011}%
	\BibitemOpen
	\bibfield  {author} {\bibinfo {author} {\bibfnamefont {C.}~\bibnamefont
			{Godr{\`e}che}},\ }\bibfield  {title} {\bibinfo {title} {Dynamics of the
			directed {{Ising}} chain},\ }\href
	{https://doi.org/10.1088/1742-5468/2011/04/P04005} {\bibfield  {journal}
		{\bibinfo  {journal} {Journal of Statistical Mechanics: Theory and
				Experiment}\ }\textbf {\bibinfo {volume} {2011}},\ \bibinfo {pages} {P04005}
		(\bibinfo {year} {2011})}\BibitemShut {NoStop}%
	\bibitem [{\citenamefont {Seara}\ \emph {et~al.}(2023)\citenamefont {Seara},
		\citenamefont {Piya},\ and\ \citenamefont
		{Tabatabai}}]{searaNonreciprocalInteractionsSpatially2023}%
	\BibitemOpen
	\bibfield  {author} {\bibinfo {author} {\bibfnamefont {D.~S.}\ \bibnamefont
			{Seara}}, \bibinfo {author} {\bibfnamefont {A.}~\bibnamefont {Piya}},\ and\
		\bibinfo {author} {\bibfnamefont {A.~P.}\ \bibnamefont {Tabatabai}},\
	}\bibfield  {title} {\bibinfo {title} {Non-reciprocal interactions spatially
			propagate fluctuations in a {{2D Ising}} model},\ }\href
	{https://doi.org/10.1088/1742-5468/accce7} {\bibfield  {journal} {\bibinfo
			{journal} {Journal of Statistical Mechanics: Theory and Experiment}\ }\textbf
		{\bibinfo {volume} {2023}},\ \bibinfo {pages} {043209} (\bibinfo {year}
		{2023})}\BibitemShut {NoStop}%
	\bibitem [{\citenamefont {Veenstra}\ \emph {et~al.}(2024)\citenamefont
		{Veenstra}, \citenamefont {Gamayun}, \citenamefont {Guo}, \citenamefont
		{Sarvi}, \citenamefont {Meinersen},\ and\ \citenamefont
		{Coulais}}]{veenstraNonreciprocalTopologicalSolitons2024}%
	\BibitemOpen
	\bibfield  {author} {\bibinfo {author} {\bibfnamefont {J.}~\bibnamefont
			{Veenstra}}, \bibinfo {author} {\bibfnamefont {O.}~\bibnamefont {Gamayun}},
		\bibinfo {author} {\bibfnamefont {X.}~\bibnamefont {Guo}}, \bibinfo {author}
		{\bibfnamefont {A.}~\bibnamefont {Sarvi}}, \bibinfo {author} {\bibfnamefont
			{C.~V.}\ \bibnamefont {Meinersen}},\ and\ \bibinfo {author} {\bibfnamefont
			{C.}~\bibnamefont {Coulais}},\ }\bibfield  {title} {\bibinfo {title}
		{Non-reciprocal topological solitons in active metamaterials},\ }\href
	{https://doi.org/10.1038/s41586-024-07097-6} {\bibfield  {journal} {\bibinfo
			{journal} {Nature}\ }\textbf {\bibinfo {volume} {627}},\ \bibinfo {pages}
		{528} (\bibinfo {year} {2024})}\BibitemShut {NoStop}%
	\bibitem [{\citenamefont {Das}\ \emph {et~al.}(2002)\citenamefont {Das},
		\citenamefont {Rao},\ and\ \citenamefont
		{Ramaswamy}}]{dasDrivenHeisenbergMagnets2002}%
	\BibitemOpen
	\bibfield  {author} {\bibinfo {author} {\bibfnamefont {J.}~\bibnamefont
			{Das}}, \bibinfo {author} {\bibfnamefont {M.}~\bibnamefont {Rao}},\ and\
		\bibinfo {author} {\bibfnamefont {S.}~\bibnamefont {Ramaswamy}},\ }\bibfield
	{title} {\bibinfo {title} {Driven {{Heisenberg}} magnets: {{Nonequilibrium}}
			criticality, spatiotemporal chaos and control},\ }\href
	{https://doi.org/10.1209/epl/i2002-00280-2} {\bibfield  {journal} {\bibinfo
			{journal} {Europhysics Letters}\ }\textbf {\bibinfo {volume} {60}},\ \bibinfo
		{pages} {418} (\bibinfo {year} {2002})}\BibitemShut {NoStop}%
	\bibitem [{\citenamefont {Bhatt}\ \emph {et~al.}()\citenamefont {Bhatt},
		\citenamefont {Mukerjee},\ and\ \citenamefont
		{Ramaswamy}}]{bhattEmergentHydrodynamicsNonreciprocal2023}%
	\BibitemOpen
	\bibfield  {author} {\bibinfo {author} {\bibfnamefont {N.}~\bibnamefont
			{Bhatt}}, \bibinfo {author} {\bibfnamefont {S.}~\bibnamefont {Mukerjee}},\
		and\ \bibinfo {author} {\bibfnamefont {S.}~\bibnamefont {Ramaswamy}},\
	}\href@noop {} {\bibinfo {title} {Emergent hydrodynamics in a non-reciprocal
			classical isotropic magnet}},\ \Eprint {https://arxiv.org/abs/2312.16500}
	{arXiv:2312.16500} \BibitemShut {NoStop}%
	\bibitem [{\citenamefont {Hatano}\ and\ \citenamefont
		{Nelson}(1997)}]{hatanoVortexPinningNonHermitian1997a}%
	\BibitemOpen
	\bibfield  {author} {\bibinfo {author} {\bibfnamefont {N.}~\bibnamefont
			{Hatano}}\ and\ \bibinfo {author} {\bibfnamefont {D.~R.}\ \bibnamefont
			{Nelson}},\ }\bibfield  {title} {\bibinfo {title} {Vortex pinning and
			non-{{Hermitian}} quantum mechanics},\ }\href
	{https://doi.org/10.1103/PhysRevB.56.8651} {\bibfield  {journal} {\bibinfo
			{journal} {Phys. Rev. B}\ }\textbf {\bibinfo {volume} {56}},\ \bibinfo
		{pages} {8651} (\bibinfo {year} {1997})}\BibitemShut {NoStop}%
	\bibitem [{\citenamefont {Hatano}\ and\ \citenamefont
		{Nelson}(1996)}]{hatanoLocalizationTransitionsNonHermitian1996a}%
	\BibitemOpen
	\bibfield  {author} {\bibinfo {author} {\bibfnamefont {N.}~\bibnamefont
			{Hatano}}\ and\ \bibinfo {author} {\bibfnamefont {D.~R.}\ \bibnamefont
			{Nelson}},\ }\bibfield  {title} {\bibinfo {title} {Localization
			{{Transitions}} in {{Non-Hermitian Quantum Mechanics}}},\ }\href
	{https://doi.org/10.1103/PhysRevLett.77.570} {\bibfield  {journal} {\bibinfo
			{journal} {Phys. Rev. Lett.}\ }\textbf {\bibinfo {volume} {77}},\ \bibinfo
		{pages} {570} (\bibinfo {year} {1996})}\BibitemShut {NoStop}%
	\bibitem [{\citenamefont {Yao}\ and\ \citenamefont
		{Wang}(2018)}]{yaoEdgeStatesTopological2018a}%
	\BibitemOpen
	\bibfield  {author} {\bibinfo {author} {\bibfnamefont {S.}~\bibnamefont
			{Yao}}\ and\ \bibinfo {author} {\bibfnamefont {Z.}~\bibnamefont {Wang}},\
	}\bibfield  {title} {\bibinfo {title} {Edge {{States}} and {{Topological
					Invariants}} of {{Non-Hermitian Systems}}},\ }\href
	{https://doi.org/10.1103/PhysRevLett.121.086803} {\bibfield  {journal}
		{\bibinfo  {journal} {Phys. Rev. Lett.}\ }\textbf {\bibinfo {volume} {121}},\
		\bibinfo {pages} {086803} (\bibinfo {year} {2018})}\BibitemShut {NoStop}%
	\bibitem [{\citenamefont {Zhang}\ \emph
		{et~al.}(2022{\natexlab{a}})\citenamefont {Zhang}, \citenamefont {Zhang},
		\citenamefont {Lu},\ and\ \citenamefont
		{Chen}}]{zhangReviewNonHermitianSkin2022}%
	\BibitemOpen
	\bibfield  {author} {\bibinfo {author} {\bibfnamefont {X.}~\bibnamefont
			{Zhang}}, \bibinfo {author} {\bibfnamefont {T.}~\bibnamefont {Zhang}},
		\bibinfo {author} {\bibfnamefont {M.-H.}\ \bibnamefont {Lu}},\ and\ \bibinfo
		{author} {\bibfnamefont {Y.-F.}\ \bibnamefont {Chen}},\ }\bibfield  {title}
	{\bibinfo {title} {A review on non-{{Hermitian}} skin effect},\ }\href
	{https://doi.org/10.1080/23746149.2022.2109431} {\bibfield  {journal}
		{\bibinfo  {journal} {Adv. Phys. X}\ }\textbf {\bibinfo {volume} {7}},\
		\bibinfo {pages} {2109431} (\bibinfo {year}
		{2022}{\natexlab{a}})}\BibitemShut {NoStop}%
	\bibitem [{\citenamefont {Okuma}\ and\ \citenamefont
		{Sato}(2023)}]{okumaNonHermitianTopologicalPhenomena2023a}%
	\BibitemOpen
	\bibfield  {author} {\bibinfo {author} {\bibfnamefont {N.}~\bibnamefont
			{Okuma}}\ and\ \bibinfo {author} {\bibfnamefont {M.}~\bibnamefont {Sato}},\
	}\bibfield  {title} {\bibinfo {title} {Non-{{Hermitian Topological
					Phenomena}}: {{A Review}}},\ }\href
	{https://doi.org/10.1146/annurev-conmatphys-040521-033133} {\bibfield
		{journal} {\bibinfo  {journal} {Annu. Rev. Condens. Matter Phys.}\ }\textbf
		{\bibinfo {volume} {14}},\ \bibinfo {pages} {83} (\bibinfo {year}
		{2023})}\BibitemShut {NoStop}%
	\bibitem [{\citenamefont {Ghosh}\ \emph {et~al.}(2024)\citenamefont {Ghosh},
		\citenamefont {Sengupta},\ and\ \citenamefont
		{Paul}}]{ghoshHilbertSpaceFragmentation2024}%
	\BibitemOpen
	\bibfield  {author} {\bibinfo {author} {\bibfnamefont {S.}~\bibnamefont
			{Ghosh}}, \bibinfo {author} {\bibfnamefont {K.}~\bibnamefont {Sengupta}},\
		and\ \bibinfo {author} {\bibfnamefont {I.}~\bibnamefont {Paul}},\ }\bibfield
	{title} {\bibinfo {title} {Hilbert space fragmentation imposed real spectrum
			of non-{{Hermitian}} systems},\ }\href
	{https://doi.org/10.1103/PhysRevB.109.045145} {\bibfield  {journal} {\bibinfo
			{journal} {Phys. Rev. B}\ }\textbf {\bibinfo {volume} {109}},\ \bibinfo
		{pages} {045145} (\bibinfo {year} {2024})}\BibitemShut {NoStop}%
	\bibitem [{\citenamefont {Many~Manda}\ \emph {et~al.}(2024)\citenamefont
		{Many~Manda}, \citenamefont {{Carretero-Gonz{\'a}lez}}, \citenamefont
		{Kevrekidis},\ and\ \citenamefont
		{Achilleos}}]{manymandaSkinModesNonlinear2024}%
	\BibitemOpen
	\bibfield  {author} {\bibinfo {author} {\bibfnamefont {B.}~\bibnamefont
			{Many~Manda}}, \bibinfo {author} {\bibfnamefont {R.}~\bibnamefont
			{{Carretero-Gonz{\'a}lez}}}, \bibinfo {author} {\bibfnamefont {P.~G.}\
			\bibnamefont {Kevrekidis}},\ and\ \bibinfo {author} {\bibfnamefont
			{V.}~\bibnamefont {Achilleos}},\ }\bibfield  {title} {\bibinfo {title} {Skin
			modes in a nonlinear {{Hatano-Nelson}} model},\ }\href
	{https://doi.org/10.1103/PhysRevB.109.094308} {\bibfield  {journal} {\bibinfo
			{journal} {Phys. Rev. B}\ }\textbf {\bibinfo {volume} {109}},\ \bibinfo
		{pages} {094308} (\bibinfo {year} {2024})}\BibitemShut {NoStop}%
	\bibitem [{\citenamefont {Zhang}\ \emph
		{et~al.}(2022{\natexlab{b}})\citenamefont {Zhang}, \citenamefont {Denner},
		\citenamefont {Bzdu{\v s}ek}, \citenamefont {Sentef},\ and\ \citenamefont
		{Neupert}}]{zhangSymmetryBreakingSpectral2022a}%
	\BibitemOpen
	\bibfield  {author} {\bibinfo {author} {\bibfnamefont {S.-B.}\ \bibnamefont
			{Zhang}}, \bibinfo {author} {\bibfnamefont {M.~M.}\ \bibnamefont {Denner}},
		\bibinfo {author} {\bibfnamefont {T.}~\bibnamefont {Bzdu{\v s}ek}}, \bibinfo
		{author} {\bibfnamefont {M.~A.}\ \bibnamefont {Sentef}},\ and\ \bibinfo
		{author} {\bibfnamefont {T.}~\bibnamefont {Neupert}},\ }\bibfield  {title}
	{\bibinfo {title} {Symmetry breaking and spectral structure of the
			interacting {{Hatano-Nelson}} model},\ }\href
	{https://doi.org/10.1103/PhysRevB.106.L121102} {\bibfield  {journal}
		{\bibinfo  {journal} {Phys. Rev. B}\ }\textbf {\bibinfo {volume} {106}},\
		\bibinfo {pages} {L121102} (\bibinfo {year}
		{2022}{\natexlab{b}})}\BibitemShut {NoStop}%
	\bibitem [{\citenamefont
		{Ezawa}(2022)}]{ezawaDynamicalNonlinearHigherorder2022a}%
	\BibitemOpen
	\bibfield  {author} {\bibinfo {author} {\bibfnamefont {M.}~\bibnamefont
			{Ezawa}},\ }\bibfield  {title} {\bibinfo {title} {Dynamical nonlinear
			higher-order non-{{Hermitian}} skin effects and topological trap-skin
			phase},\ }\href {https://doi.org/10.1103/PhysRevB.105.125421} {\bibfield
		{journal} {\bibinfo  {journal} {Phys. Rev. B}\ }\textbf {\bibinfo {volume}
			{105}},\ \bibinfo {pages} {125421} (\bibinfo {year} {2022})}\BibitemShut
	{NoStop}%
	\bibitem [{\citenamefont {Zhu}\ \emph {et~al.}(2022)\citenamefont {Zhu},
		\citenamefont {Wang}, \citenamefont {Leykam}, \citenamefont {Xue},
		\citenamefont {Wang},\ and\ \citenamefont
		{Chong}}]{zhuAnomalousSingleModeLasing2022}%
	\BibitemOpen
	\bibfield  {author} {\bibinfo {author} {\bibfnamefont {B.}~\bibnamefont
			{Zhu}}, \bibinfo {author} {\bibfnamefont {Q.}~\bibnamefont {Wang}}, \bibinfo
		{author} {\bibfnamefont {D.}~\bibnamefont {Leykam}}, \bibinfo {author}
		{\bibfnamefont {H.}~\bibnamefont {Xue}}, \bibinfo {author} {\bibfnamefont
			{Q.~J.}\ \bibnamefont {Wang}},\ and\ \bibinfo {author} {\bibfnamefont
			{Y.~D.}\ \bibnamefont {Chong}},\ }\bibfield  {title} {\bibinfo {title}
		{Anomalous {{Single-Mode Lasing Induced}} by {{Nonlinearity}} and the
			{{Non-Hermitian Skin Effect}}},\ }\href
	{https://doi.org/10.1103/PhysRevLett.129.013903} {\bibfield  {journal}
		{\bibinfo  {journal} {Phys. Rev. Lett.}\ }\textbf {\bibinfo {volume} {129}},\
		\bibinfo {pages} {013903} (\bibinfo {year} {2022})}\BibitemShut {NoStop}%
	\bibitem [{\citenamefont {Brighi}\ and\ \citenamefont
		{Nunnenkamp}(2024)}]{Brighi2024}%
	\BibitemOpen
	\bibfield  {author} {\bibinfo {author} {\bibfnamefont {P.}~\bibnamefont
			{Brighi}}\ and\ \bibinfo {author} {\bibfnamefont {A.}~\bibnamefont
			{Nunnenkamp}},\ }\bibfield  {title} {\bibinfo {title} {Nonreciprocal dynamics
			and the non-hermitian skin effect of repulsively bound pairs},\ }\href
	{https://doi.org/10.1103/PhysRevA.110.L020201} {\bibfield  {journal}
		{\bibinfo  {journal} {Phys. Rev. A}\ }\textbf {\bibinfo {volume} {110}},\
		\bibinfo {pages} {L020201} (\bibinfo {year} {2024})}\BibitemShut {NoStop}%
	\bibitem [{\citenamefont
		{Longhi}(2025)}]{longhiModulationalInstabilityDynamical2025}%
	\BibitemOpen
	\bibfield  {author} {\bibinfo {author} {\bibfnamefont {S.}~\bibnamefont
			{Longhi}},\ }\bibfield  {title} {\bibinfo {title} {Modulational
			{{Instability}} and {{Dynamical Growth Blockade}} in the {{Nonlinear
					Hatano}}--{{Nelson Model}}},\ }\href {https://doi.org/10.1002/apxr.202400154}
	{\bibfield  {journal} {\bibinfo  {journal} {Adv. Phys. Res.}\ }\textbf
		{\bibinfo {volume} {4}},\ \bibinfo {pages} {2400154} (\bibinfo {year}
		{2025})}\BibitemShut {NoStop}%
	\bibitem [{\citenamefont {Kasprzak}\ \emph {et~al.}(2006)\citenamefont
		{Kasprzak}, \citenamefont {Richard}, \citenamefont {Kundermann},
		\citenamefont {Baas}, \citenamefont {Jeambrun}, \citenamefont {Keeling},
		\citenamefont {Marchetti}, \citenamefont {Szyma{\'n}ska}, \citenamefont
		{Andr{\'e}}, \citenamefont {Staehli}, \citenamefont {Savona}, \citenamefont
		{Littlewood}, \citenamefont {Deveaud},\ and\ \citenamefont
		{Dang}}]{kasprzakBoseEinsteinCondensation2006}%
	\BibitemOpen
	\bibfield  {author} {\bibinfo {author} {\bibfnamefont {J.}~\bibnamefont
			{Kasprzak}}, \bibinfo {author} {\bibfnamefont {M.}~\bibnamefont {Richard}},
		\bibinfo {author} {\bibfnamefont {S.}~\bibnamefont {Kundermann}}, \bibinfo
		{author} {\bibfnamefont {A.}~\bibnamefont {Baas}}, \bibinfo {author}
		{\bibfnamefont {P.}~\bibnamefont {Jeambrun}}, \bibinfo {author}
		{\bibfnamefont {J.~M.~J.}\ \bibnamefont {Keeling}}, \bibinfo {author}
		{\bibfnamefont {F.~M.}\ \bibnamefont {Marchetti}}, \bibinfo {author}
		{\bibfnamefont {M.~H.}\ \bibnamefont {Szyma{\'n}ska}}, \bibinfo {author}
		{\bibfnamefont {R.}~\bibnamefont {Andr{\'e}}}, \bibinfo {author}
		{\bibfnamefont {J.~L.}\ \bibnamefont {Staehli}}, \bibinfo {author}
		{\bibfnamefont {V.}~\bibnamefont {Savona}}, \bibinfo {author} {\bibfnamefont
			{P.~B.}\ \bibnamefont {Littlewood}}, \bibinfo {author} {\bibfnamefont
			{B.}~\bibnamefont {Deveaud}},\ and\ \bibinfo {author} {\bibfnamefont {L.~S.}\
			\bibnamefont {Dang}},\ }\bibfield  {title} {\bibinfo {title}
		{Bose--{{Einstein}} condensation of exciton polaritons},\ }\href
	{https://doi.org/10.1038/nature05131} {\bibfield  {journal} {\bibinfo
			{journal} {Nature}\ }\textbf {\bibinfo {volume} {443}},\ \bibinfo {pages}
		{409} (\bibinfo {year} {2006})}\BibitemShut {NoStop}%
	\bibitem [{\citenamefont {Amo}\ \emph {et~al.}(2009)\citenamefont {Amo},
		\citenamefont {Sanvitto}, \citenamefont {Laussy}, \citenamefont {Ballarini},
		\citenamefont {Valle}, \citenamefont {Martin}, \citenamefont {Lema{\^i}tre},
		\citenamefont {Bloch}, \citenamefont {Krizhanovskii}, \citenamefont
		{Skolnick}, \citenamefont {Tejedor},\ and\ \citenamefont
		{Vi{\~n}a}}]{amoCollectiveFluidDynamics2009}%
	\BibitemOpen
	\bibfield  {author} {\bibinfo {author} {\bibfnamefont {A.}~\bibnamefont
			{Amo}}, \bibinfo {author} {\bibfnamefont {D.}~\bibnamefont {Sanvitto}},
		\bibinfo {author} {\bibfnamefont {F.~P.}\ \bibnamefont {Laussy}}, \bibinfo
		{author} {\bibfnamefont {D.}~\bibnamefont {Ballarini}}, \bibinfo {author}
		{\bibfnamefont {E.~D.}\ \bibnamefont {Valle}}, \bibinfo {author}
		{\bibfnamefont {M.~D.}\ \bibnamefont {Martin}}, \bibinfo {author}
		{\bibfnamefont {A.}~\bibnamefont {Lema{\^i}tre}}, \bibinfo {author}
		{\bibfnamefont {J.}~\bibnamefont {Bloch}}, \bibinfo {author} {\bibfnamefont
			{D.~N.}\ \bibnamefont {Krizhanovskii}}, \bibinfo {author} {\bibfnamefont
			{M.~S.}\ \bibnamefont {Skolnick}}, \bibinfo {author} {\bibfnamefont
			{C.}~\bibnamefont {Tejedor}},\ and\ \bibinfo {author} {\bibfnamefont
			{L.}~\bibnamefont {Vi{\~n}a}},\ }\bibfield  {title} {\bibinfo {title}
		{Collective fluid dynamics of a polariton condensate in a semiconductor
			microcavity},\ }\href {https://doi.org/10.1038/nature07640} {\bibfield
		{journal} {\bibinfo  {journal} {Nature}\ }\textbf {\bibinfo {volume} {457}},\
		\bibinfo {pages} {291} (\bibinfo {year} {2009})}\BibitemShut {NoStop}%
	\bibitem [{\citenamefont {Klaers}\ \emph {et~al.}(2010)\citenamefont {Klaers},
		\citenamefont {Schmitt}, \citenamefont {Vewinger},\ and\ \citenamefont
		{Weitz}}]{klaersBoseEinsteinCondensation2010}%
	\BibitemOpen
	\bibfield  {author} {\bibinfo {author} {\bibfnamefont {J.}~\bibnamefont
			{Klaers}}, \bibinfo {author} {\bibfnamefont {J.}~\bibnamefont {Schmitt}},
		\bibinfo {author} {\bibfnamefont {F.}~\bibnamefont {Vewinger}},\ and\
		\bibinfo {author} {\bibfnamefont {M.}~\bibnamefont {Weitz}},\ }\bibfield
	{title} {\bibinfo {title} {Bose--{{Einstein}} condensation of photons in an
			optical microcavity},\ }\href {https://doi.org/10.1038/nature09567}
	{\bibfield  {journal} {\bibinfo  {journal} {Nature}\ }\textbf {\bibinfo
			{volume} {468}},\ \bibinfo {pages} {545} (\bibinfo {year}
		{2010})}\BibitemShut {NoStop}%
	\bibitem [{\citenamefont {Wertz}\ \emph {et~al.}(2012)\citenamefont {Wertz},
		\citenamefont {Amo}, \citenamefont {Solnyshkov}, \citenamefont {Ferrier},
		\citenamefont {Liew}, \citenamefont {Sanvitto}, \citenamefont {Senellart},
		\citenamefont {Sagnes}, \citenamefont {Lema{\^i}tre}, \citenamefont
		{Kavokin}, \citenamefont {Malpuech},\ and\ \citenamefont
		{Bloch}}]{wertzPropagationAmplificationDynamics2012}%
	\BibitemOpen
	\bibfield  {author} {\bibinfo {author} {\bibfnamefont {E.}~\bibnamefont
			{Wertz}}, \bibinfo {author} {\bibfnamefont {A.}~\bibnamefont {Amo}}, \bibinfo
		{author} {\bibfnamefont {D.~D.}\ \bibnamefont {Solnyshkov}}, \bibinfo
		{author} {\bibfnamefont {L.}~\bibnamefont {Ferrier}}, \bibinfo {author}
		{\bibfnamefont {T.~C.~H.}\ \bibnamefont {Liew}}, \bibinfo {author}
		{\bibfnamefont {D.}~\bibnamefont {Sanvitto}}, \bibinfo {author}
		{\bibfnamefont {P.}~\bibnamefont {Senellart}}, \bibinfo {author}
		{\bibfnamefont {I.}~\bibnamefont {Sagnes}}, \bibinfo {author} {\bibfnamefont
			{A.}~\bibnamefont {Lema{\^i}tre}}, \bibinfo {author} {\bibfnamefont {A.~V.}\
			\bibnamefont {Kavokin}}, \bibinfo {author} {\bibfnamefont {G.}~\bibnamefont
			{Malpuech}},\ and\ \bibinfo {author} {\bibfnamefont {J.}~\bibnamefont
			{Bloch}},\ }\bibfield  {title} {\bibinfo {title} {Propagation and
			{{Amplification Dynamics}} of {{1D Polariton Condensates}}},\ }\href
	{https://doi.org/10.1103/PhysRevLett.109.216404} {\bibfield  {journal}
		{\bibinfo  {journal} {Phys. Rev. Lett.}\ }\textbf {\bibinfo {volume} {109}},\
		\bibinfo {pages} {216404} (\bibinfo {year} {2012})}\BibitemShut {NoStop}%
	\bibitem [{\citenamefont {Marcos}\ \emph {et~al.}(2012)\citenamefont {Marcos},
		\citenamefont {Tomadin}, \citenamefont {Diehl},\ and\ \citenamefont
		{Rabl}}]{marcosPhotonCondensationCircuit2012}%
	\BibitemOpen
	\bibfield  {author} {\bibinfo {author} {\bibfnamefont {D.}~\bibnamefont
			{Marcos}}, \bibinfo {author} {\bibfnamefont {A.}~\bibnamefont {Tomadin}},
		\bibinfo {author} {\bibfnamefont {S.}~\bibnamefont {Diehl}},\ and\ \bibinfo
		{author} {\bibfnamefont {P.}~\bibnamefont {Rabl}},\ }\bibfield  {title}
	{\bibinfo {title} {Photon condensation in circuit quantum electrodynamics by
			engineered dissipation},\ }\href
	{https://doi.org/10.1088/1367-2630/14/5/055005} {\bibfield  {journal}
		{\bibinfo  {journal} {New J. Phys.}\ }\textbf {\bibinfo {volume} {14}},\
		\bibinfo {pages} {055005} (\bibinfo {year} {2012})}\BibitemShut {NoStop}%
	\bibitem [{\citenamefont {He}\ \emph {et~al.}(2015)\citenamefont {He},
		\citenamefont {Sieberer}, \citenamefont {Altman},\ and\ \citenamefont
		{Diehl}}]{heScalingPropertiesOnedimensional2015}%
	\BibitemOpen
	\bibfield  {author} {\bibinfo {author} {\bibfnamefont {L.}~\bibnamefont
			{He}}, \bibinfo {author} {\bibfnamefont {L.~M.}\ \bibnamefont {Sieberer}},
		\bibinfo {author} {\bibfnamefont {E.}~\bibnamefont {Altman}},\ and\ \bibinfo
		{author} {\bibfnamefont {S.}~\bibnamefont {Diehl}},\ }\bibfield  {title}
	{\bibinfo {title} {Scaling properties of one-dimensional driven-dissipative
			condensates},\ }\href {https://doi.org/10.1103/PhysRevB.92.155307} {\bibfield
		{journal} {\bibinfo  {journal} {Phys. Rev. B}\ }\textbf {\bibinfo {volume}
			{92}},\ \bibinfo {pages} {155307} (\bibinfo {year} {2015})}\BibitemShut
	{NoStop}%
	\bibitem [{\citenamefont {Keeling}\ and\ \citenamefont
		{Berloff}(2008)}]{keelingSpontaneousRotatingVortex2008}%
	\BibitemOpen
	\bibfield  {author} {\bibinfo {author} {\bibfnamefont {J.}~\bibnamefont
			{Keeling}}\ and\ \bibinfo {author} {\bibfnamefont {N.~G.}\ \bibnamefont
			{Berloff}},\ }\bibfield  {title} {\bibinfo {title} {Spontaneous {{Rotating
					Vortex Lattices}} in a {{Pumped Decaying Condensate}}},\ }\href
	{https://doi.org/10.1103/PhysRevLett.100.250401} {\bibfield  {journal}
		{\bibinfo  {journal} {Phys. Rev. Lett.}\ }\textbf {\bibinfo {volume} {100}},\
		\bibinfo {pages} {250401} (\bibinfo {year} {2008})}\BibitemShut {NoStop}%
	\bibitem [{\citenamefont {Marino}\ and\ \citenamefont
		{Diehl}(2016)}]{marinoDrivenMarkovianQuantum2016a}%
	\BibitemOpen
	\bibfield  {author} {\bibinfo {author} {\bibfnamefont {J.}~\bibnamefont
			{Marino}}\ and\ \bibinfo {author} {\bibfnamefont {S.}~\bibnamefont {Diehl}},\
	}\bibfield  {title} {\bibinfo {title} {Driven {{Markovian Quantum
					Criticality}}},\ }\href {https://doi.org/10.1103/PhysRevLett.116.070407}
	{\bibfield  {journal} {\bibinfo  {journal} {Phys. Rev. Lett.}\ }\textbf
		{\bibinfo {volume} {116}},\ \bibinfo {pages} {070407} (\bibinfo {year}
		{2016})}\BibitemShut {NoStop}%
	\bibitem [{\citenamefont {He}\ \emph {et~al.}(2017)\citenamefont {He},
		\citenamefont {Sieberer},\ and\ \citenamefont
		{Diehl}}]{heSpaceTimeVortexDriven2017}%
	\BibitemOpen
	\bibfield  {author} {\bibinfo {author} {\bibfnamefont {L.}~\bibnamefont
			{He}}, \bibinfo {author} {\bibfnamefont {L.~M.}\ \bibnamefont {Sieberer}},\
		and\ \bibinfo {author} {\bibfnamefont {S.}~\bibnamefont {Diehl}},\ }\bibfield
	{title} {\bibinfo {title} {Space-{{Time Vortex Driven Crossover}} and
			{{Vortex Turbulence Phase Transition}} in {{One-Dimensional Driven Open
					Condensates}}},\ }\href {https://doi.org/10.1103/PhysRevLett.118.085301}
	{\bibfield  {journal} {\bibinfo  {journal} {Phys. Rev. Lett.}\ }\textbf
		{\bibinfo {volume} {118}},\ \bibinfo {pages} {085301} (\bibinfo {year}
		{2017})}\BibitemShut {NoStop}%
	\bibitem [{\citenamefont {Vercesi}\ \emph {et~al.}(2023)\citenamefont
		{Vercesi}, \citenamefont {Fontaine}, \citenamefont {Ravets}, \citenamefont
		{Bloch}, \citenamefont {Richard}, \citenamefont {Canet},\ and\ \citenamefont
		{Minguzzi}}]{vercesiPhaseDiagramOnedimensional2023}%
	\BibitemOpen
	\bibfield  {author} {\bibinfo {author} {\bibfnamefont {F.}~\bibnamefont
			{Vercesi}}, \bibinfo {author} {\bibfnamefont {Q.}~\bibnamefont {Fontaine}},
		\bibinfo {author} {\bibfnamefont {S.}~\bibnamefont {Ravets}}, \bibinfo
		{author} {\bibfnamefont {J.}~\bibnamefont {Bloch}}, \bibinfo {author}
		{\bibfnamefont {M.}~\bibnamefont {Richard}}, \bibinfo {author} {\bibfnamefont
			{L.}~\bibnamefont {Canet}},\ and\ \bibinfo {author} {\bibfnamefont
			{A.}~\bibnamefont {Minguzzi}},\ }\bibfield  {title} {\bibinfo {title} {Phase
			diagram of one-dimensional driven-dissipative exciton-polariton
			condensates},\ }\href {https://doi.org/10.1103/PhysRevResearch.5.043062}
	{\bibfield  {journal} {\bibinfo  {journal} {Phys. Rev. Res.}\ }\textbf
		{\bibinfo {volume} {5}},\ \bibinfo {pages} {043062} (\bibinfo {year}
		{2023})}\BibitemShut {NoStop}%
	\bibitem [{\citenamefont {Fontaine}\ \emph {et~al.}(2022)\citenamefont
		{Fontaine}, \citenamefont {Squizzato}, \citenamefont {Baboux}, \citenamefont
		{Amelio}, \citenamefont {Lema{\^i}tre}, \citenamefont {Morassi},
		\citenamefont {Sagnes}, \citenamefont {Le~Gratiet}, \citenamefont {Harouri},
		\citenamefont {Wouters}, \citenamefont {Carusotto}, \citenamefont {Amo},
		\citenamefont {Richard}, \citenamefont {Minguzzi}, \citenamefont {Canet}
		\emph {et~al.}}]{fontaineKardarParisiZhang2022}%
	\BibitemOpen
	\bibfield  {author} {\bibinfo {author} {\bibfnamefont {Q.}~\bibnamefont
			{Fontaine}}, \bibinfo {author} {\bibfnamefont {D.}~\bibnamefont {Squizzato}},
		\bibinfo {author} {\bibfnamefont {F.}~\bibnamefont {Baboux}}, \bibinfo
		{author} {\bibfnamefont {I.}~\bibnamefont {Amelio}}, \bibinfo {author}
		{\bibfnamefont {A.}~\bibnamefont {Lema{\^i}tre}}, \bibinfo {author}
		{\bibfnamefont {M.}~\bibnamefont {Morassi}}, \bibinfo {author} {\bibfnamefont
			{I.}~\bibnamefont {Sagnes}}, \bibinfo {author} {\bibfnamefont
			{L.}~\bibnamefont {Le~Gratiet}}, \bibinfo {author} {\bibfnamefont
			{A.}~\bibnamefont {Harouri}}, \bibinfo {author} {\bibfnamefont
			{M.}~\bibnamefont {Wouters}}, \bibinfo {author} {\bibfnamefont
			{I.}~\bibnamefont {Carusotto}}, \bibinfo {author} {\bibfnamefont
			{A.}~\bibnamefont {Amo}}, \bibinfo {author} {\bibfnamefont {M.}~\bibnamefont
			{Richard}}, \bibinfo {author} {\bibfnamefont {A.}~\bibnamefont {Minguzzi}},
		\bibinfo {author} {\bibfnamefont {L.}~\bibnamefont {Canet}}, \emph {et~al.},\
	}\bibfield  {title} {\bibinfo {title} {Kardar--{{Parisi}}--{{Zhang}}
			universality in a one-dimensional polariton condensate},\ }\href
	{https://doi.org/10.1038/s41586-022-05001-8} {\bibfield  {journal} {\bibinfo
			{journal} {Nature}\ }\textbf {\bibinfo {volume} {608}},\ \bibinfo {pages}
		{687} (\bibinfo {year} {2022})}\BibitemShut {NoStop}%
	\bibitem [{\citenamefont {Mandal}\ \emph {et~al.}(2020)\citenamefont {Mandal},
		\citenamefont {Banerjee}, \citenamefont {Ostrovskaya},\ and\ \citenamefont
		{Liew}}]{mandalNonreciprocalTransportExciton2020}%
	\BibitemOpen
	\bibfield  {author} {\bibinfo {author} {\bibfnamefont {S.}~\bibnamefont
			{Mandal}}, \bibinfo {author} {\bibfnamefont {R.}~\bibnamefont {Banerjee}},
		\bibinfo {author} {\bibfnamefont {E.~A.}\ \bibnamefont {Ostrovskaya}},\ and\
		\bibinfo {author} {\bibfnamefont {T.~C.~H.}\ \bibnamefont {Liew}},\
	}\bibfield  {title} {\bibinfo {title} {Nonreciprocal {{Transport}} of
			{{Exciton Polaritons}} in a {{Non-Hermitian Chain}}},\ }\href
	{https://doi.org/10.1103/PhysRevLett.125.123902} {\bibfield  {journal}
		{\bibinfo  {journal} {Phys. Rev. Lett.}\ }\textbf {\bibinfo {volume} {125}},\
		\bibinfo {pages} {123902} (\bibinfo {year} {2020})}\BibitemShut {NoStop}%
	\bibitem [{\citenamefont {Tao}\ \emph {et~al.}()\citenamefont {Tao},
		\citenamefont {{Mercado-Gutierrez}}, \citenamefont {Zhao},\ and\
		\citenamefont {Spielman}}]{taoImaginaryGaugePotentials2025}%
	\BibitemOpen
	\bibfield  {author} {\bibinfo {author} {\bibfnamefont {J.}~\bibnamefont
			{Tao}}, \bibinfo {author} {\bibfnamefont {E.}~\bibnamefont
			{{Mercado-Gutierrez}}}, \bibinfo {author} {\bibfnamefont {M.}~\bibnamefont
			{Zhao}},\ and\ \bibinfo {author} {\bibfnamefont {I.}~\bibnamefont
			{Spielman}},\ }\bibfield  {title} {\bibinfo {title} {Imaginary gauge
			potentials in a non-{{Hermitian}} spin-orbit coupled quantum gas},\
	}\href@noop {} {\ }\Eprint {https://arxiv.org/abs/2504.08614}
	{arXiv:2504.08614} \BibitemShut {NoStop}%
	\bibitem [{\citenamefont {Hanai}\ \emph {et~al.}(2019)\citenamefont {Hanai},
		\citenamefont {Edelman}, \citenamefont {Ohashi},\ and\ \citenamefont
		{Littlewood}}]{Hanai2019polariton}%
	\BibitemOpen
	\bibfield  {author} {\bibinfo {author} {\bibfnamefont {R.}~\bibnamefont
			{Hanai}}, \bibinfo {author} {\bibfnamefont {A.}~\bibnamefont {Edelman}},
		\bibinfo {author} {\bibfnamefont {Y.}~\bibnamefont {Ohashi}},\ and\ \bibinfo
		{author} {\bibfnamefont {P.~B.}\ \bibnamefont {Littlewood}},\ }\bibfield
	{title} {\bibinfo {title} {Non-hermitian phase transition from a polariton
			bose-einstein condensate to a photon laser},\ }\href
	{https://doi.org/10.1103/PhysRevLett.122.185301} {\bibfield  {journal}
		{\bibinfo  {journal} {Phys. Rev. Lett.}\ }\textbf {\bibinfo {volume} {122}},\
		\bibinfo {pages} {185301} (\bibinfo {year} {2019})}\BibitemShut {NoStop}%
	\bibitem [{\citenamefont {Hanai}\ and\ \citenamefont
		{Littlewood}(2020)}]{Hanai2020CEP}%
	\BibitemOpen
	\bibfield  {author} {\bibinfo {author} {\bibfnamefont {R.}~\bibnamefont
			{Hanai}}\ and\ \bibinfo {author} {\bibfnamefont {P.~B.}\ \bibnamefont
			{Littlewood}},\ }\bibfield  {title} {\bibinfo {title} {Critical fluctuations
			at a many-body exceptional point},\ }\href
	{https://doi.org/10.1103/PhysRevResearch.2.033018} {\bibfield  {journal}
		{\bibinfo  {journal} {Phys. Rev. Res.}\ }\textbf {\bibinfo {volume} {2}},\
		\bibinfo {pages} {033018} (\bibinfo {year} {2020})}\BibitemShut {NoStop}%
	\bibitem [{\citenamefont {Suchanek}\ \emph {et~al.}(2023)\citenamefont
		{Suchanek}, \citenamefont {Kroy},\ and\ \citenamefont
		{Loos}}]{Suchanek2023CEPentropyprod}%
	\BibitemOpen
	\bibfield  {author} {\bibinfo {author} {\bibfnamefont {T.}~\bibnamefont
			{Suchanek}}, \bibinfo {author} {\bibfnamefont {K.}~\bibnamefont {Kroy}},\
		and\ \bibinfo {author} {\bibfnamefont {S.~A.~M.}\ \bibnamefont {Loos}},\
	}\bibfield  {title} {\bibinfo {title} {Irreversible mesoscale fluctuations
			herald the emergence of dynamical phases},\ }\href
	{https://doi.org/10.1103/PhysRevLett.131.258302} {\bibfield  {journal}
		{\bibinfo  {journal} {Phys. Rev. Lett.}\ }\textbf {\bibinfo {volume} {131}},\
		\bibinfo {pages} {258302} (\bibinfo {year} {2023})}\BibitemShut {NoStop}%
	\bibitem [{\citenamefont {Nakanishi}\ \emph {et~al.}(2024)\citenamefont
		{Nakanishi}, \citenamefont {Hanai},\ and\ \citenamefont
		{Sasamoto}}]{nakanishi2024continuoustimecrystalspt}%
	\BibitemOpen
	\bibfield  {author} {\bibinfo {author} {\bibfnamefont {Y.}~\bibnamefont
			{Nakanishi}}, \bibinfo {author} {\bibfnamefont {R.}~\bibnamefont {Hanai}},\
		and\ \bibinfo {author} {\bibfnamefont {T.}~\bibnamefont {Sasamoto}},\
	}\href@noop {} {\bibinfo {title} {Continuous time crystals as a pt symmetric
			state and the emergence of critical exceptional points}} (\bibinfo {year}
	{2024}),\ \Eprint {https://arxiv.org/abs/2406.09018} {arXiv:2406.09018}
	\BibitemShut {NoStop}%
	\bibitem [{\citenamefont {Weis}\ \emph {et~al.}()\citenamefont {Weis},
		\citenamefont {Fruchart}, \citenamefont {Hanai}, \citenamefont {Kawagoe},
		\citenamefont {Littlewood},\ and\ \citenamefont
		{Vitelli}}]{weisExceptionalPointsNonlinear2023}%
	\BibitemOpen
	\bibfield  {author} {\bibinfo {author} {\bibfnamefont {C.}~\bibnamefont
			{Weis}}, \bibinfo {author} {\bibfnamefont {M.}~\bibnamefont {Fruchart}},
		\bibinfo {author} {\bibfnamefont {R.}~\bibnamefont {Hanai}}, \bibinfo
		{author} {\bibfnamefont {K.}~\bibnamefont {Kawagoe}}, \bibinfo {author}
		{\bibfnamefont {P.~B.}\ \bibnamefont {Littlewood}},\ and\ \bibinfo {author}
		{\bibfnamefont {V.}~\bibnamefont {Vitelli}},\ }\bibfield  {title} {\bibinfo
		{title} {Exceptional points in nonlinear and stochastic dynamics},\
	}\href@noop {} {\ }\Eprint {https://arxiv.org/abs/2207.11667}
	{arXiv:2207.11667} \BibitemShut {NoStop}%
	\bibitem [{\citenamefont {Breuer}\ and\ \citenamefont
		{Petruccione}(2010)}]{breuerTheoryOpenQuantum2010}%
	\BibitemOpen
	\bibfield  {author} {\bibinfo {author} {\bibfnamefont {H.-P.}\ \bibnamefont
			{Breuer}}\ and\ \bibinfo {author} {\bibfnamefont {F.}~\bibnamefont
			{Petruccione}},\ }\href@noop {} {\emph {\bibinfo {title} {The Theory of Open
				Quantum Systems}}},\ \bibinfo {edition} {repr}\ ed.\ (\bibinfo  {publisher}
	{Clarendon Press},\ \bibinfo {address} {Oxford},\ \bibinfo {year}
	{2010})\BibitemShut {NoStop}%
	\bibitem [{\citenamefont {Metelmann}\ and\ \citenamefont
		{T{\"u}reci}(2018)}]{metelmannNonreciprocalSignalRouting2018}%
	\BibitemOpen
	\bibfield  {author} {\bibinfo {author} {\bibfnamefont {A.}~\bibnamefont
			{Metelmann}}\ and\ \bibinfo {author} {\bibfnamefont {H.~E.}\ \bibnamefont
			{T{\"u}reci}},\ }\bibfield  {title} {\bibinfo {title} {Nonreciprocal signal
			routing in an active quantum network},\ }\href
	{https://doi.org/10.1103/PhysRevA.97.043833} {\bibfield  {journal} {\bibinfo
			{journal} {Phys. Rev. A}\ }\textbf {\bibinfo {volume} {97}},\ \bibinfo
		{pages} {043833} (\bibinfo {year} {2018})}\BibitemShut {NoStop}%
	\bibitem [{\citenamefont {Wanjura}\ \emph {et~al.}(2020)\citenamefont
		{Wanjura}, \citenamefont {Brunelli},\ and\ \citenamefont
		{Nunnenkamp}}]{wanjuraTopologicalFrameworkDirectional2020a}%
	\BibitemOpen
	\bibfield  {author} {\bibinfo {author} {\bibfnamefont {C.~C.}\ \bibnamefont
			{Wanjura}}, \bibinfo {author} {\bibfnamefont {M.}~\bibnamefont {Brunelli}},\
		and\ \bibinfo {author} {\bibfnamefont {A.}~\bibnamefont {Nunnenkamp}},\
	}\bibfield  {title} {\bibinfo {title} {Topological framework for directional
			amplification in driven-dissipative cavity arrays},\ }\href
	{https://doi.org/10.1038/s41467-020-16863-9} {\bibfield  {journal} {\bibinfo
			{journal} {Nat Commun}\ }\textbf {\bibinfo {volume} {11}},\ \bibinfo {pages}
		{3149} (\bibinfo {year} {2020})}\BibitemShut {NoStop}%
	\bibitem [{\citenamefont {McDonald}\ \emph {et~al.}(2022)\citenamefont
		{McDonald}, \citenamefont {Hanai},\ and\ \citenamefont
		{Clerk}}]{mcdonaldNonequilibriumStationaryStates2022}%
	\BibitemOpen
	\bibfield  {author} {\bibinfo {author} {\bibfnamefont {A.}~\bibnamefont
			{McDonald}}, \bibinfo {author} {\bibfnamefont {R.}~\bibnamefont {Hanai}},\
		and\ \bibinfo {author} {\bibfnamefont {A.~A.}\ \bibnamefont {Clerk}},\
	}\bibfield  {title} {\bibinfo {title} {Nonequilibrium stationary states of
			quantum non-{{Hermitian}} lattice models},\ }\href
	{https://doi.org/10.1103/PhysRevB.105.064302} {\bibfield  {journal} {\bibinfo
			{journal} {Phys. Rev. B}\ }\textbf {\bibinfo {volume} {105}},\ \bibinfo
		{pages} {064302} (\bibinfo {year} {2022})}\BibitemShut {NoStop}%
	\bibitem [{sup()}]{supp}%
	\BibitemOpen
	\href@noop {} {}\bibinfo {note} {See Supplemental Material for additional
		details concerning the stability of the PBC solutions, the spectrum of
		Hatano-Nelson model, and additional details concerning the dynamical phases
		for OBC.}\BibitemShut {Stop}%
	\bibitem [{\citenamefont {Aranson}\ and\ \citenamefont
		{Kramer}(2002)}]{aransonWorldComplexGinzburgLandau2002b}%
	\BibitemOpen
	\bibfield  {author} {\bibinfo {author} {\bibfnamefont {I.~S.}\ \bibnamefont
			{Aranson}}\ and\ \bibinfo {author} {\bibfnamefont {L.}~\bibnamefont
			{Kramer}},\ }\bibfield  {title} {\bibinfo {title} {The world of the complex
			{{Ginzburg-Landau}} equation},\ }\href
	{https://doi.org/10.1103/RevModPhys.74.99} {\bibfield  {journal} {\bibinfo
			{journal} {Rev. Mod. Phys.}\ }\textbf {\bibinfo {volume} {74}},\ \bibinfo
		{pages} {99} (\bibinfo {year} {2002})}\BibitemShut {NoStop}%
	\bibitem [{\citenamefont {Ravoux}\ \emph {et~al.}(2000)\citenamefont {Ravoux},
		\citenamefont {Le~Diz{\`e}s},\ and\ \citenamefont
		{Le~Gal}}]{ravouxStabilityAnalysisPlane2000}%
	\BibitemOpen
	\bibfield  {author} {\bibinfo {author} {\bibfnamefont {J.~F.}\ \bibnamefont
			{Ravoux}}, \bibinfo {author} {\bibfnamefont {S.}~\bibnamefont
			{Le~Diz{\`e}s}},\ and\ \bibinfo {author} {\bibfnamefont {P.}~\bibnamefont
			{Le~Gal}},\ }\bibfield  {title} {\bibinfo {title} {Stability analysis of
			plane wave solutions of the discrete {{Ginzburg-Landau}} equation},\ }\href
	{https://doi.org/10.1103/PhysRevE.61.390} {\bibfield  {journal} {\bibinfo
			{journal} {Phys. Rev. E}\ }\textbf {\bibinfo {volume} {61}},\ \bibinfo
		{pages} {390} (\bibinfo {year} {2000})}\BibitemShut {NoStop}%
	\bibitem [{\citenamefont {Kawabata}\ \emph {et~al.}(2019)\citenamefont
		{Kawabata}, \citenamefont {Shiozaki}, \citenamefont {Ueda},\ and\
		\citenamefont {Sato}}]{kawabataSymmetryTopologyNonHermitian2019}%
	\BibitemOpen
	\bibfield  {author} {\bibinfo {author} {\bibfnamefont {K.}~\bibnamefont
			{Kawabata}}, \bibinfo {author} {\bibfnamefont {K.}~\bibnamefont {Shiozaki}},
		\bibinfo {author} {\bibfnamefont {M.}~\bibnamefont {Ueda}},\ and\ \bibinfo
		{author} {\bibfnamefont {M.}~\bibnamefont {Sato}},\ }\bibfield  {title}
	{\bibinfo {title} {Symmetry and {{Topology}} in {{Non-Hermitian Physics}}},\
	}\href {https://doi.org/10.1103/PhysRevX.9.041015} {\bibfield  {journal}
		{\bibinfo  {journal} {Phys. Rev. X}\ }\textbf {\bibinfo {volume} {9}},\
		\bibinfo {pages} {041015} (\bibinfo {year} {2019})}\BibitemShut {NoStop}%
	\bibitem [{\citenamefont {Iemini}\ \emph {et~al.}(2018)\citenamefont {Iemini},
		\citenamefont {Russomanno}, \citenamefont {Keeling}, \citenamefont
		{Schir{\`o}}, \citenamefont {Dalmonte},\ and\ \citenamefont
		{Fazio}}]{ieminiBoundaryTimeCrystals2018}%
	\BibitemOpen
	\bibfield  {author} {\bibinfo {author} {\bibfnamefont {F.}~\bibnamefont
			{Iemini}}, \bibinfo {author} {\bibfnamefont {A.}~\bibnamefont {Russomanno}},
		\bibinfo {author} {\bibfnamefont {J.}~\bibnamefont {Keeling}}, \bibinfo
		{author} {\bibfnamefont {M.}~\bibnamefont {Schir{\`o}}}, \bibinfo {author}
		{\bibfnamefont {M.}~\bibnamefont {Dalmonte}},\ and\ \bibinfo {author}
		{\bibfnamefont {R.}~\bibnamefont {Fazio}},\ }\bibfield  {title} {\bibinfo
		{title} {Boundary {{Time Crystals}}},\ }\href
	{https://doi.org/10.1103/PhysRevLett.121.035301} {\bibfield  {journal}
		{\bibinfo  {journal} {Phys. Rev. Lett.}\ }\textbf {\bibinfo {volume} {121}},\
		\bibinfo {pages} {035301} (\bibinfo {year} {2018})}\BibitemShut {NoStop}%
	\bibitem [{\citenamefont {Khemani}\ \emph {et~al.}(2019)\citenamefont
		{Khemani}, \citenamefont {Moessner},\ and\ \citenamefont
		{Sondhi}}]{khemaniBriefHistoryTime2019}%
	\BibitemOpen
	\bibfield  {author} {\bibinfo {author} {\bibfnamefont {V.}~\bibnamefont
			{Khemani}}, \bibinfo {author} {\bibfnamefont {R.}~\bibnamefont {Moessner}},\
		and\ \bibinfo {author} {\bibfnamefont {S.~L.}\ \bibnamefont {Sondhi}},\
	}\bibfield  {title} {\bibinfo {title} {A {{Brief History}} of {{Time
					Crystals}}},\ }\href@noop {} {\  (\bibinfo {year} {2019})},\ \Eprint
	{https://arxiv.org/abs/1910.10745} {arXiv:1910.10745} \BibitemShut {NoStop}%
	\bibitem [{\citenamefont {Benettin}\ \emph {et~al.}(1980)\citenamefont
		{Benettin}, \citenamefont {Galgani}, \citenamefont {Giorgilli},\ and\
		\citenamefont {Strelcyn}}]{benettinLyapunovCharacteristicExponents1980}%
	\BibitemOpen
	\bibfield  {author} {\bibinfo {author} {\bibfnamefont {G.}~\bibnamefont
			{Benettin}}, \bibinfo {author} {\bibfnamefont {L.}~\bibnamefont {Galgani}},
		\bibinfo {author} {\bibfnamefont {A.}~\bibnamefont {Giorgilli}},\ and\
		\bibinfo {author} {\bibfnamefont {J.-M.}\ \bibnamefont {Strelcyn}},\
	}\bibfield  {title} {\bibinfo {title} {Lyapunov {{Characteristic Exponents}}
			for smooth dynamical systems and for hamiltonian systems; {{A}} method for
			computing all of them. {{Part}} 2: {{Numerical}} application},\ }\href
	{https://doi.org/10.1007/BF02128237} {\bibfield  {journal} {\bibinfo
			{journal} {Meccanica}\ }\textbf {\bibinfo {volume} {15}},\ \bibinfo {pages}
		{21} (\bibinfo {year} {1980})}\BibitemShut {NoStop}%
	\bibitem [{\citenamefont {Rossler}(1979)}]{rosslerEquationHyperchaos1979}%
	\BibitemOpen
	\bibfield  {author} {\bibinfo {author} {\bibfnamefont {O.~E.}\ \bibnamefont
			{Rossler}},\ }\bibfield  {title} {\bibinfo {title} {An equation for
			hyperchaos},\ }\href {https://doi.org/10.1016/0375-9601(79)90150-6}
	{\bibfield  {journal} {\bibinfo  {journal} {Physics Letters A}\ }\textbf
		{\bibinfo {volume} {71}},\ \bibinfo {pages} {155} (\bibinfo {year}
		{1979})}\BibitemShut {NoStop}%
	\bibitem [{\citenamefont {R{\"o}ssler}(1991)}]{rosslerChaoticHierarchy1991}%
	\BibitemOpen
	\bibfield  {author} {\bibinfo {author} {\bibfnamefont {O.~E.}\ \bibnamefont
			{R{\"o}ssler}},\ }\bibfield  {title} {\bibinfo {title} {The chaotic
			hierarchy},\ }in\ \href {https://doi.org/10.1142/9789814503372_0002} {\emph
		{\bibinfo {booktitle} {A {{Chaotic Hierarchy}}}}}\ (\bibinfo  {publisher}
	{WORLD SCIENTIFIC},\ \bibinfo {year} {1991})\ pp.\ \bibinfo {pages}
	{25--47}\BibitemShut {NoStop}%
\end{thebibliography}
\end{document}


\title{Supplemental Material for:
Phase Transitions in Nonreciprocal Driven-Dissipative Condensates}
\date{\today}
\author{Ron Belyansky}
\affiliation{Pritzker School of Molecular Engineering, University of Chicago, Chicago, IL 60637, USA}
\author{Cheyne Weis}
\affiliation{James Franck Institute and Department of Physics,
The University of Chicago, Chicago, Illinois 60637, United States}
\author{Ryo Hanai}
\affiliation{Department of Physics, Institute of Science Tokyo, 2-12-1 Ookayama Meguro-ku, Tokyo, 152-8551, Japan}
\author{Peter B. Littlewood}
\affiliation{James Franck Institute and Department of Physics,
The University of Chicago, Chicago, Illinois 60637, United States}
\affiliation{School of Physics and Astronomy, University of St Andrews, N Haugh, St Andrews KY16 9SS, United Kingdom
}
\author{Aashish A. Clerk}
\affiliation{Pritzker School of Molecular Engineering, University of Chicago, Chicago, IL 60637, USA}
\maketitle
\onecolumngrid

\renewcommand{\theequation}{S\arabic{equation}}

\renewcommand{\bibnumfmt}[1]{[S#1]}
\renewcommand{\citenumfont}[1]{S#1} 

\pagenumbering{arabic}

\makeatletter
\renewcommand{\thefigure}{S\@arabic\c@figure}
\renewcommand \thetable{S\@arabic\c@table}

This Supplemental Material is organized as follows. In \cref{sec:PBC}, we derive the stability condition of the condensate solutions for the PBC system. In \cref{sec:HN-spectrum}, we derive the spectrum of the OBC Hatano-Nelson, including Eq. (5) of the main text. In \cref{sec:OBC}, we provide additional details on the phases and phases transitions in the OBC system. In particular, in \cref{sec:CEP} we demonstrate the existence of critical exceptional points; in \cref{sec:chaos}, give details
of the chaotic dynamics contained in the model; in Sec. \cref{sec:edge}, we discuss the edge dynamics throughout the dynamic
portion of the phase diagram.

\section{Periodic boundary conditions}
\label{sec:PBC}

In this section, we derive the stability condition of the PBC solutions from Eq. (4) of the main text that lead to the plot shown in Fig. 2 of the main text.

The momentum-q condenstate solution for PBC is given by
\begin{equation}
\alpha_j(t) = r_q e^{iqj-i\omega_qt}\qc r_q = \sqrt{\frac{\kappa-\gamma_q}{\Gamma}}\qc \gamma_q = 2\gamma(1+\sin(q+\theta)) \qc \omega_q = -2J\cos(q), 
\end{equation}
for all $q = 2\pi m/N,\, m= 0,\cdots, N-1$ satisfying $\kappa -\gamma_q >0$.

We consider momentum-space fluctuations on top of the momentum-q condensate of the form
\begin{equation}
\alpha_j(t) = r_q e^{-i\omega_q t+iqj} +e^{-i\omega_q t}\sum_ke^{i kj}\delta\alpha_k.
\end{equation}

To first order in $\delta\alpha_k$, we get
\begin{equation}
\partial_t \delta\alpha_k = \qty(\kappa-\gamma_k-2\Gamma r_q^2-i(\omega_k-\omega_q))\delta\alpha_k -\Gamma r_q^2\delta\alpha^*_{2q-k} \equiv \beta_k\delta\alpha_k-\Lambda \delta\alpha^*_{2q-k}.
\end{equation}

The stability of the momentum-q condensate is then determined by the eigenvalues of 
\begin{equation}
\partial_t \mqty(\delta\alpha_{k}\\\delta\alpha^*_{2q-k}) = \mqty(\beta_k&-\Lambda\\-\Lambda&{\beta}^*_{2q-k})\mqty(\delta\alpha_{k}\\\delta\alpha^*_{2q-k}).
\end{equation}
The eigenvalues are given by
\begin{equation}
\lambda_{k,\pm} = \frac{\beta_k +\beta^*_{2q-k}}{2} \pm\frac{1}{2}\sqrt{(\beta_k-{\beta}^*_{2q-k})^2+\Lambda^2}.
\end{equation}
The momentum-$q$ traveling wave is stable if 
\begin{equation}
\label{eq:stability-cond}
\max_{k}[\Re(\lambda_{k,+})] <0.
\end{equation} 
We evaluate \cref{eq:stability-cond} numerically to produce Fig. 2 of the main text.
While a full analytical expression for \cref{eq:stability-cond} is too cumbersome, we can obtain a simple result in the limit of $\kappa\rightarrow \infty$. In that case, we obtain $\lambda_{k,+} =-2\gamma (\cos(k)-1)\sin(q+\theta)$.
The stability condition in this limit thus becomes 
\begin{equation}
    \sin(q+\theta)<0.
\end{equation}
For $\theta=\pi$, we see that only right-moving modes, $0<q<\pi$, are stable as $\kappa\rightarrow\infty$, consistent with the numerical results in Fig. 2 of the main text.

\section{Hatano-Nelson}
\label{sec:HN-spectrum}

In this section, we derive the spectrum of the OBC Hatano-Nelson, shown in Eq.~(5) of the main text.
The Hatano-Nelson Hamiltonian [Eq.~(1) of the main text] is
\begin{equation}
	\label{eq:HN_OBC_matrix}
H_{\mathrm{HN}}=\mqty(\Delta&J_+\\
J_-&\Delta&J_+\\
&J_-&\ddots&\ddots\\
&&\ddots&\ddots&).
\end{equation}
The eigenvalue equation, $H_{\mathrm{HN}}\psi = \lambda \psi$ yields to the bulk equations:
\begin{equation}
	\label{eq:bulk-eq}
J_- \psi_{j-1} +E\psi_j +J_+ \psi_{j+1} =0\qc 1<j<N,
\end{equation}
where $E\equiv \Delta -\lambda$.
For the boundaries, we have instead
\begin{align}
	\label{eq:boundaries-eq}
E \psi_1+J_+\psi_2 =0,\\
J_- \psi_{N-1}+ E \psi_N =0.
\end{align}

The bulk equation \cref{eq:bulk-eq} is solved with the eigenvector $\psi(z) = \mqty(z,z^2,\cdots,z^N)^T$ giving
\begin{equation}
	\label{eq:bulk-cond}
J_-+Ez+J_+z^2=0.
\end{equation}
For a given $E$, \cref{eq:bulk-cond} is a quadratic equation that has two solutions $z_1,z_2$. Hence the general solution is a superposition $\Psi = c_1\psi(z_1)+c_2\psi(z_2)$.
Plugging this into the boundaries equations \cref{eq:boundaries-eq} gives
\begin{equation}
H_B\mqty(c_1\\c_2)=0 \qq{where} H_B = 
\mqty(Ez_1+J_+z_1^2&Ez_2+J_+z_2^2\\J_-z_1^{N-1}+Ez_1^N&J_-z_2^{N-1}+Ez_2^N).
\end{equation}
$H_B$ has a nontrivial nullspace iff $\det(H_B)=0$, implying (assuming $J_-\neq 0$)
\begin{equation}
	\label{eq:boundary-cond-zz}
-z_1^{N+1}+z_2^{N+1}=0 \qand c_1=-c_2.
\end{equation}

For $J_-\neq 0$, \cref{eq:bulk-cond} implies
\begin{equation}
	\label{eq:z1z2}
z_1z_2 = \frac{J_-}{J_+},
\end{equation}
which we can solve by 
\begin{equation}
	\label{eq:z1z2-anstaz}
z_1 = re^{i\theta}e^{i\delta/2}\qc z_2 = re^{-i\theta}e^{i\delta/2}\qq{where} r^2 = \abs{\frac{J_-}{J_+}}\qand \arg\qty(\frac{J_-}{J_+}) = \delta.
\end{equation}
Plugging this into \cref{eq:boundary-cond-zz} yields
\begin{equation}
r^{N+1} \sin(\theta(N+1))=0 \rightarrow \theta_m =\frac{\pi m}{N+1}\qc m=1,\cdots N.
\end{equation}

\Cref{eq:bulk-cond} then gives the eigenvalues
\begin{dmath}
\lambda_m = \Delta+2e^{-i\delta/2}J_-\sqrt{\abs{\frac{J_+}{J_-}}}\cos(\theta_m),
\end{dmath}
and the corresponding eigenvectors are 
\begin{equation}
\Psi_m = \mqty(re^{i\delta/2}\sin(\theta_m),r^2e^{i\delta}\sin(2\theta_m),\cdots,r^Ne^{i\delta N/2}\sin(\theta_mN))^T.
\end{equation}

\section{Open boundary conditions}
\label{sec:OBC}
In this section, we provide additional plots and examples of some of the salient features in the model with open boundary conditions. In Section \ref{sec:CEP}, we demonstrate the occurrence of the critical exceptional point at the boundary of the Static phase and Phase I. In Section \ref{sec:chaos}, give details of the chaotic dynamics contained in the model. Finally, in Section \ref{sec:edge}, give explore the various edge behaviors exhibited by the model.
\subsection{Critical exceptional points}
\label{sec:CEP}

\begin{figure}[b!]
	\centering
	\includegraphics[width=6in]{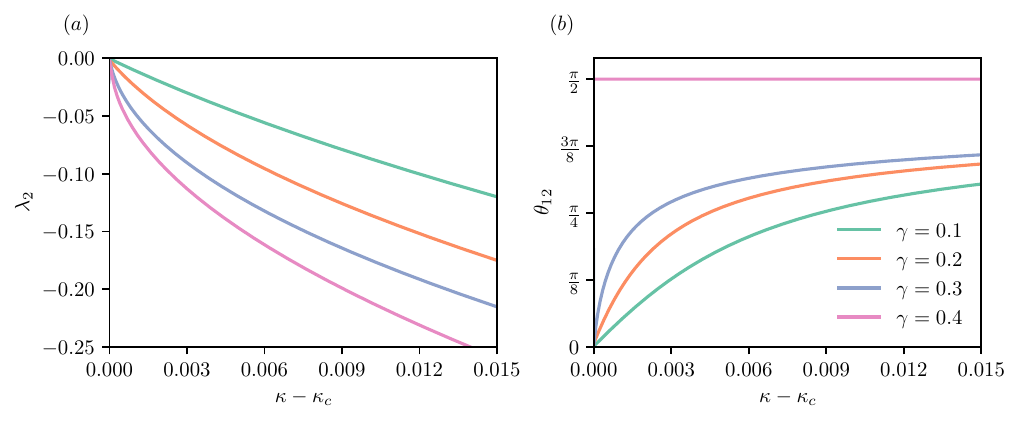}
	\caption{Demonstration of the critical exceptional point (CEP) between the vacuum and Phase I. (a) The second largest eigenvalue $\lambda_2$ of the Jacobian as a function of the distance from the critical point $\kappa- \kappa_c$. (b) The angle between the first two eigenvectors $\theta_{12}$ of the Jacobian. The critical values for $\gamma = \{0.1,0.2,0.3,0.4\}$ are respectively $\kappa_c \approx \{2.3734,2.3848,2.3893, 2.3935\}$.}
	\label{fig:EP}
\end{figure}

Transitions between static and dynamic $\mathbb{Z}_2$-symmetry broken phases have been previously shown to occur at a critical exceptional points (CEP), a critical point that has multiple zero modes with the corresponding eigenvectors aligned \cite{hanaiCriticalFluctuationsManybody2020, weisExceptionalPointsNonlinear2023, fruchartNonreciprocalPhaseTransitions2021a, zelleUniversalPhenomenologyCritical2024}. A critical exceptional line exists on the red line in Fig. 3(a) of the main text, excluding the portion forming the border between the static phase and Phase IV as shown at $\gamma =0.4$ in Fig. \ref{fig:EP}. We numerically demonstrate the CEP in Fig. \ref{fig:EP} for values of $\gamma$ which approach Phase I (i.e., $\gamma =\{0.1,0.2,0.3\}$). The dynamical matrix (i.e. the Jacobian of the equations of motion,  defined as $J_{ij} = \frac{\partial\dot{\alpha_i}}{\partial\alpha_j}$ with $i$ and $j$ indexing both the sites and their complex conjugates) is evaluated on the steady state solution for values of $\kappa$ approaching the dynamical region from the static phase. At $\kappa=\kappa_c$, the Jacobian exhibits a 2nd order exceptional point, where the mode undergoing critical slowing aligns with the existing Goldstone mode from $U(1)$ symmetry breaking. We expect the transition to be an exceptional point when approached from the limit cycle phase as well, as found in previous works \cite{fruchartNonreciprocalPhaseTransitions2021a, weisExceptionalPointsNonlinear2023}.
\subsection{Chaotic Dynamics}
\label{sec:chaos}
\begin{figure}
    \centering
	\includegraphics[width=\textwidth]{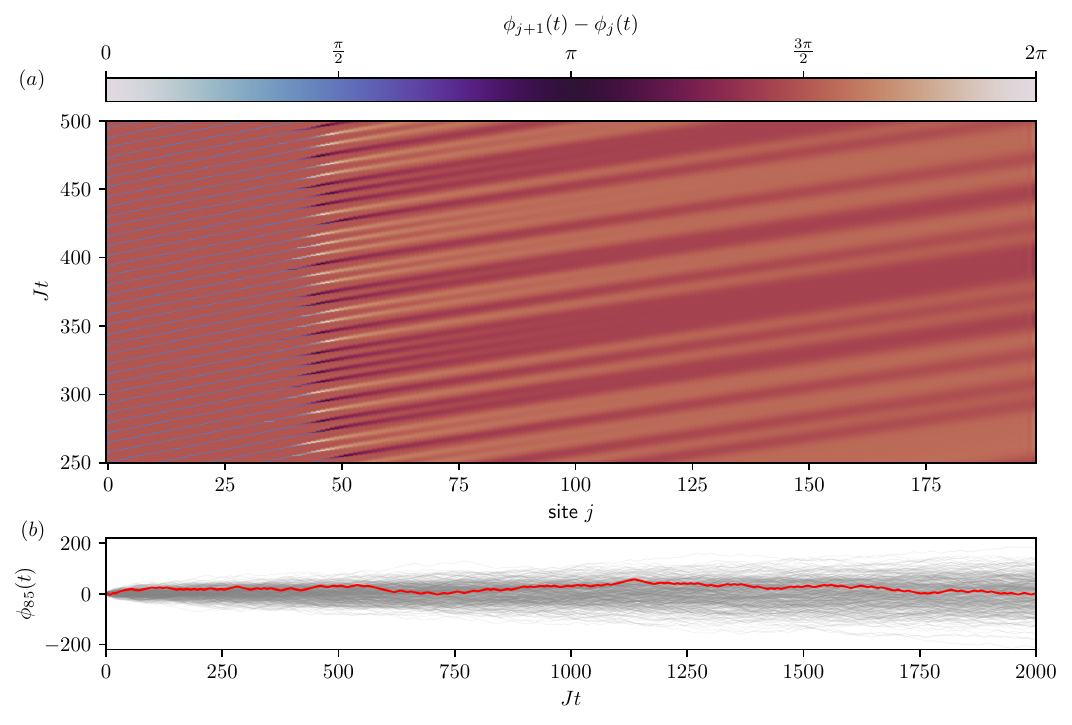}
	\caption{(a) 
    Time and space behavior of adjacent phase differences for $\kappa=2.2, \ \gamma =0.5$,  demonstrating chaotic  dynamics. (b) Temporal behavior of the phase for a site in the bulk ($j=85$), for 300 nearby trajectories on the chaotic attractor with an example trajectory highlighted in red. 
    }
    \label{fig:chaosZ2}
\end{figure}

In the chaotic/hyperchaotic portions of Phase IV, time translation symmetry is completely broken due to dynamics that never repeat in time. The system exhibits domains of clockwise (CW) and counterclockwise (CCW) rotation corresponding to the two colors of stripes in Fig. \ref{fig:chaosZ2}(a). Intermittent switching between directions of rotation leads to the diffusion of the phase variables for sites further from the edge. The trajectory exhibits no net preference for CW or CCW motion at long times, as illustrated in Fig. \ref{fig:chaosZ2}(b). The phase trajectories  with nearby initial conditions diverge from each other in a manner resembling an unbiased random walk due to the lack of preference for either direction of rotation. The random behavior in the bulk is in stark contrast to the seemingly deterministic edge behavior.

\subsection{Edge dynamics}
\label{sec:edge}
\begin{figure}[t!]
	\centering
	\includegraphics[width=\linewidth]{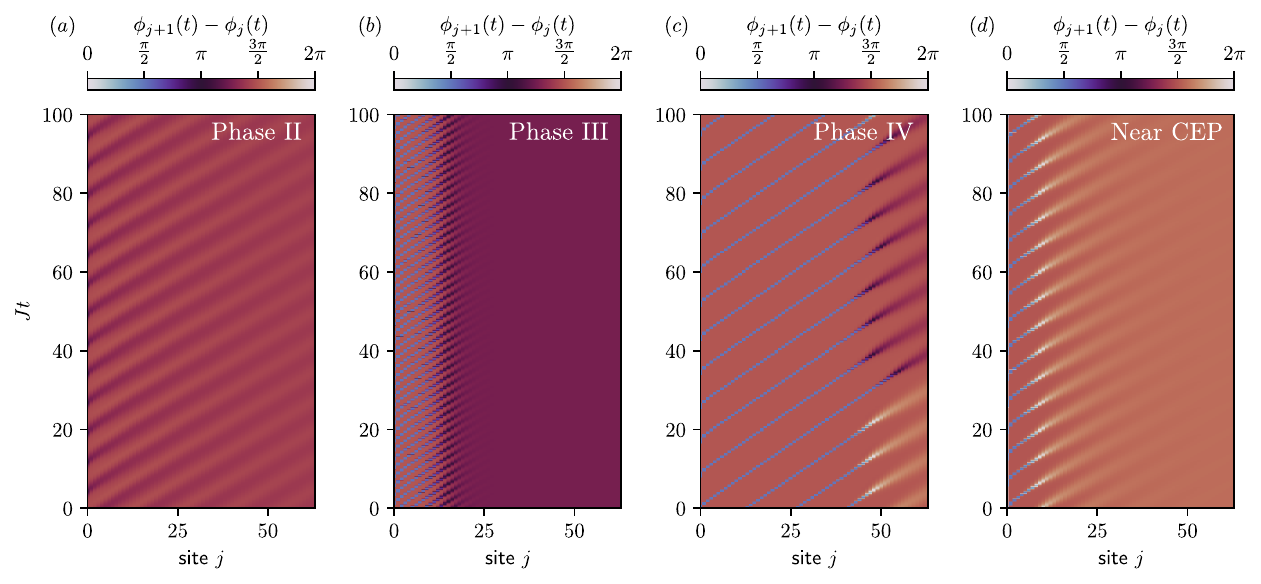}
	\caption{(a) Phase II dynamics at the point $\kappa = 2.2,\ \gamma=0.22$ (b) An example of a bistable solution with edge dynamics in Phase III for $\kappa = 1.25,\ \gamma=0.4$. (c) Close-up of the periodic edge behavior in the chaotic phase for $\kappa = 2.2,\ \gamma=0.7$. (d) Atypical dynamics near the Hatano-Nelson exceptional point at $\kappa = 2.1 ,\ \gamma=0.8$.}
    \label{fig:edge_phenomena}
\end{figure}

To conclude, we explore the irregular dynamics that can occur at the edge of the chain. One example was already demonstrated at the phase transition occurring at the boundary of Phase I and II . The bulk dynamics on either side of this transition are identical, but the state exhibits quasiperiodic behavior in Phase II, particularly near the left edge (see Fig. 5(b) of the main text). The dynamics in Phase II is shown in \cref{fig:edge_phenomena}(a), where the nonuniform phase oscillations decay into the bulk also demonstrated in the inset of Fig. 3(e) of the main text. Phase III is similarly only distinguished from Phase I by amplitude and phase oscillations occurring near the edge. A hierarchy of these steady state solutions exist, as demonstrated in Fig. 4(d) of the main text. The quantized length scale distinguishing consecutive edge modes can be found for this choice of parameters via integration of the curves in Fig. 4(d), of which the mean difference is $\sim4.2$ sites, roughly half the wavelength of the bulk phase oscillations, $\sim8.2$ sites.

Phase IV (Right), the hyperchaotic phase, displays distinct edge dynamics compared to Phase IV (Left). The first $\sim40$ sites of the chain predominantly exhibit amplitude dynamics. Domain walls separating regions of $\phi = 0$ and $\phi= \pi$ are created at the edge and propagated into the bulk. These solitons continue until, at a seemingly arbitrary distance into the bulk, they are converted into boundaries between domains of CW and CCW rotation. This iterative domain wall generation results in the random walk in the phase of the bulk sites, as described above.

Approaching the Hatano-Nelson exceptional point from the dynamic phase, additional atypical dynamics can be exhibited beyond the features discussed thus far. As demonstrated in \ref{fig:edge_phenomena}(d), the edge dynamics mixes together the periodic domain walls of the chaotic Phase IV (Right) and the nonuniform phase oscillations of Phase II.

